\documentclass[journal]{IEEEtran}
\usepackage[]{graphicx}
\usepackage{amsmath}
\usepackage{amssymb}
\usepackage{algorithmic}
\usepackage{algorithm}
\usepackage{caption}
\usepackage{mathtools}
\usepackage{changes}
\usepackage{multirow}
\usepackage{verbatim}

\ifCLASSINFOpdf
\else
\fi
\begin{document}
\title{Frequency-Limited Pseudo-Optimal Rational Krylov Algorithm for Power System Reduction}
\author{Umair~Zulfiqar,~Victor~Sreeram, and Xin~Du
\thanks{U.~Zulfiqar and V.~Sreeram are with the School of Electrical, Electronics and Computer Engineering, The University of Western Australia, 35 Stirling Highway, Crawley, WA 6009 (email: umair.zulfiqar@research.uwa.edu.au, victor.sreeram@uwa.edu.au).}
\thanks{X.~Du is with the School of Mechatronic Engineering and Automation, Shanghai University, Shanghai 200072, China, and also with the Shanghai Key Laboratory of Power Station Automation Technology, Shanghai University, Shanghai 200444, P.R. China  (e-mail: duxin@shu.edu.cn)}}
\markboth{\tiny{This paper is accepted for publication in International Journal of Electrical Power \& Energy Systems: DOI: 10.1016/j.ijepes.2019.105798}. Copyright reserved by Elsevier.}
{Shell \MakeLowercase{\textit{et al.}}: Bare Demo of IEEEtran.cls for IEEE Journals}
\maketitle
\begin{abstract}
In this paper, a computationally efficient frequency-limited model reduction algorithm is presented for large-scale interconnected power systems. The algorithm generates a reduced order model which not only preserves the electromechanical modes of the original power system but also satisfies a subset of the first-order optimality conditions for $\mathcal{H}_{2,\omega}$ model reduction problem within the desired frequency interval. The reduced order model accurately captures the oscillatory behavior of the original power system and provides a good time- and frequency-domain accuracy. The proposed algorithm enables fast simulation, analysis, and damping controller design for the original large-scale power system. The efficacy of the proposed algorithm is validated on benchmark power system examples.
\end{abstract}
\begin{IEEEkeywords}
Electromechanical modes, Krylov subspace, Modal preservation, Model reduction, Power system oscillations.
\end{IEEEkeywords}
\IEEEpeerreviewmaketitle

\section{Introduction}
\begin{figure}[!h]
\centering
\includegraphics[width=8.5cm]{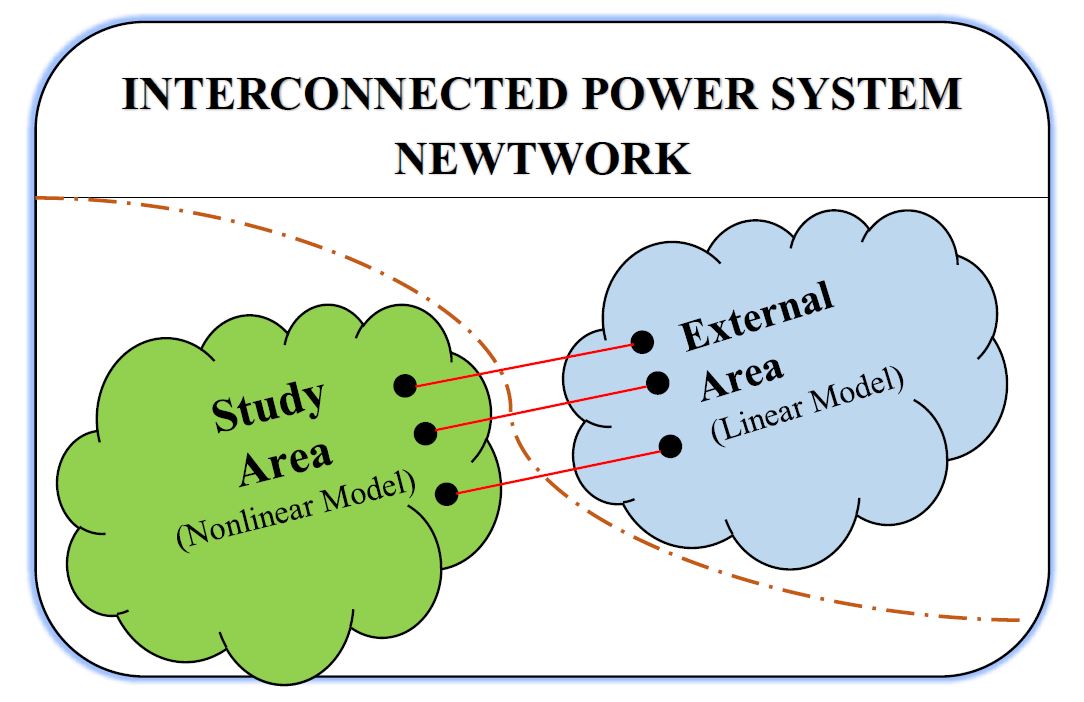}
\caption{Partitioning of power system for MOR}\label{fig:0}
\end{figure}

\IEEEPARstart{T}{oday's} power system is a large network of interconnected power apparatus like generators, lines, and buses that covers a large geographical territory. There is a growing trend to facilitate further interconnections with neighboring systems, and thus the size of the interconnected power system network is likely to continue to increase. The mathematical representation of these large-scale power system networks can easily reach several thousands of differential equations. This poses a challenge for fast and efficient simulation, analysis, and control system design for these large-scale power systems despite a significant growth in the storage and computational capabilities in recent time \cite{sturk2014coherency}. Model order reduction (MOR) offers a solution to the problem by providing a reduced order model (ROM), which enables fast simulation and control system design without significantly affecting the accuracy. MOR is generally referred to as ``dynamic equivalency" in the power system literature \cite{chow2013power}.\\
The analysis of a complete power system network with every subtle detail is neither practical nor required. In the MOR of power systems, the power system is first partitioned according to the importance \cite{sturk2014coherency}\nocite{huang2012model}\nocite{chow2013power}-\nocite{scarciotti2015model}\nocite{chaniotis2005model}\cite{nath1985coherency}. The portion of the power system under investigation, which contains the important variables, constitutes the study area, and it is mathematically described by a detailed nonlinear model. Note that this study area is not reduced. On the other hand, the portion whose effect on the analysis in the study area is only of the interest is mathematically described by a linear model, and it constitutes the external area; see Fig. \ref{fig:0}. MOR is applied to the linear model of the external area. For instance, not only that a linear model suffice in the small-signal stability analysis and damping controller design, it can be further reduced using MOR techniques without a significant loss of accuracy \cite{pal2006robust}.
\\The coherency-based MOR methods have been historically employed to obtain a dynamically equivalent ROM \cite{nath1985coherency}-\nocite{podmore1978identification}\cite{de1975coherency}. The response of coherent generators is similar to a particular set of inputs. The first step in the coherency-based MOR techniques is to identify and group the coherent set of generators and construct a lumped system. A ROM is then obtained from the lumped model by exploiting the physical properties of electrical machines connected to the power system network. The dependence on physical properties restricts the flexible applicability of these methods. Recently, an increasing interest in MOR techniques which rely on the mathematical properties instead of the physical properties of the power system apparatus is shown by the power system community \cite{scarciotti2015model,scarciotti2017low}. For instance, balanced truncation and moment matching have been successfully used in power system reduction, showing some promising results \cite{ghosh2013balanced}-\nocite{zhu2016power}\cite{freitas2008gramian}.\\
The power systems exhibit local and interarea oscillations in the frequency region between $0.8-2$ Hz and $0.1-0.7$ Hz, respectively \cite{kundur1994power}. These are associated with the poorly damped modes of the power system model and are often called ``electromechanical or critical modes". These modes are crucial for small-signal stability analysis and for the damping controller design. Therefore, these modes must be preserved in the ROM to retain the oscillatory behavior of the original model. The frequency response of the ROM should closely match that of the original system within $0.1-2$ Hz. The importance of good frequency-domain accuracy within $0.1-2$ Hz has been recognized consistently in the literature; see for instance \cite{chaniotis2005model,sanchez1996power}. In \cite{chaniotis2005model}, the ROM interpolates the original system at and around zero frequency to effectively capture these oscillations in the frequency-domain. In \cite{sanchez1996power}, it is suggested to retain the critical modes in the ROM to preserve the oscillation associated with these modes.\\
The preservation of slow and poorly damped modes in the ROM is beneficial from the damping controller design perspective \cite{sanchez1996power}, and it also improves the accuracy of the ROM in the time-domain \cite{scarciotti2015model,scarciotti2017low,yogarathinam2017new}. Most of the MOR algorithms used for power system reduction like balanced truncation \cite{moore1981principal} and moment matching \cite{chaniotis2005model} do not have modal preservation property. It is customary to increase the order of ROM in these algorithms in a hope to capture the poorly damped critical modes of the original system in the ROM. This popular belief has recently been refuted in \cite{scarciotti2015model}, and it is argued that there is no guarantee to capture these modes in the ROM by increasing the order. It is further shown that the quality of the ROM can be improved by preserving the slow and poorly damped modes instead of increasing its order \cite{scarciotti2015model}. In \cite{yogarathinam2017new}, an $\mathcal{H}_2$-MOR algorithm is proposed for power systems that includes modal preservation as a cost function of its optimality criteria. The algorithm does preserve the electromechanical modes in the ROM, but the first-order optimality conditions (as defined in \cite{wilson1970optimum,gugercin2008h_2}) of $\mathcal{H}_2$-MOR are no longer satisfied with this heuristic modification in \cite{gugercin2008h_2}. It gives good frequency and time domain accuracy, but the excessive computational cost associated with the particle swarm optimization technique \cite{kennedy2010particle} makes it unsuitable for large-scale systems. In \cite{zulfiqar2019finite}, the power system reduction is considered as a finite-frequency MOR problem with an additional constraint that the electromechanical modes of the systems are preserved in the ROM. The algorithm is computationally efficient, but the ROM of acceptable accuracy is not that compact because it uses modal truncation to preserve critical modes. The order of ROM should be significantly larger than the number of modes to be preserved. The accuracy in the specified frequency region is obtained by using frequency-dependent extended realization of the original system. MOR is applied to this extended realization, and the ROM is obtained via an inverse transformation. In \cite{petersson2014model}, the optimal frequency-limited $\mathcal{H}_2$-MOR problem is considered, and an algorithm is proposed, which generates an optimal ROM. The algorithm requires the solution of Lyapunov equations and linear matrix inequalities (LMIs) to find the optimal ROM, which is not feasible in a large-scale setting. In \cite{vuillemin2014frequency}, the problem is described as bi-tangential Hermite interpolation, which can be solved in a computationally efficient way. However, the original system is required to be converted into pole-residue form, which is again computationally not feasible in a large-scale setting. Moreover, both the algorithm \cite{petersson2014model} and \cite{vuillemin2014frequency} are iterative algorithms with no guarantee on the convergence, and they do not have a modal preservation property.\\
In this paper, we consider the same problem of \cite{zulfiqar2019finite} and propose a computationally efficient MOR algorithm that ensures a good accuracy in the specified frequency region with explicit modal preservation. Unlike \cite{zulfiqar2019finite}, a fairly compact ROM can be obtained using the proposed algorithm, and the order of ROM can even be equal to the number of modes to be preserved. The algorithm uses a moment matching approach based on the parametrized family of ROM \cite{ahmad2011krylov} and generates a ROM which satisfies a subset of the first-order optimality conditions for frequency-limited $\mathcal{H}_2$-MOR problem \cite{vuillemin2014frequency}. Unlike \cite{petersson2014model} and \cite{vuillemin2014frequency}, the proposed algorithm is iteration-free and does not requires the solutions of large-scale Lyapunov equations, LMIs, and pole-residue form. The performance of the proposed algorithm is tested by considering benchmark power system reduction problems.
\section{Preliminaries}
Consider an $e_n$-machine, $k$-bus system as the external area which is connected to $p$-buses of the study area via $p$-tie lines. The external area can be described by the following second-order classical model \cite{liu2013krylov} used for power system reduction for $i=1,\cdots,e_n$, i.e.,
\begin{align}
\dot{\bar{\delta}}_i&=\bar{\omega}_i-\bar{\omega}_s\nonumber\\
\frac{2\bar{H}_i}{\bar{\omega}_s}\dot{\bar{\omega}}_i&=\bar{T}_i-\bar{D_i}(\bar{\omega}_i-\bar{\omega}_s)\nonumber\\
-\bar{E}_i&\sum_{j=1}^{e_n}\big(\bar{E}_jG_{ij}cos(\bar{\delta}_i-\bar{\delta}_j)+\bar{E}_jB_{ij}sin(\bar{\delta}_i-\bar{\delta}_j)\big)\nonumber\\
-\bar{E}_i&\sum_{j=1}^{p}\big(\bar{V}_j\bar{G}_{ij}cos(\bar{\delta}_i-\bar{\theta}_j)+\bar{V}_j\bar{B}_{ij}sin(\bar{\delta}_i-\bar{\theta}_j)\big).\label{eq:37a}
\end{align}
$\bar{H}_i$, $\bar{D}_i$, $\bar{\delta}_i$, $\bar{\omega}_i$, $\bar{E}_i$, and $\bar{T}_i$ are the inertial coefficient, damping coefficient, rotor angle, angular velocity, internal voltage, and mechanical input power respectively of the machine $i$ of the external area. $\bar{\omega}_s$ is the reference angular velocity. $\bar{V}_j$ and $\bar{\theta}_j$ are the voltage magnitude and angle on the $p$-buses of the study area. The admittance $G_{ij}+jB_{ij}$ connects machine $i$ with machine $j$, and the admittance $\bar{G}_{ij}+j\bar{B}_{ij}$ connects machine $i$ with the boundary bus $j$.
The nonlinear model in equation (\ref{eq:37a}) can be linearized around an equilibrium point to obtain a $n^{th}$ order state-space model, i.e.,
\begin{align}
\begin{bmatrix}\Delta \dot{\bar{\delta}}\\\Delta \dot{\bar{\omega}}\end{bmatrix}=A\begin{bmatrix}\Delta\bar{\delta}\\ \Delta\bar{\omega}\end{bmatrix}+B\begin{bmatrix}\Delta\bar{\theta}_j\\\Delta\bar{V}_j\end{bmatrix}, && \bar{\delta}_j=C\begin{bmatrix}\Delta\bar{\delta}\\\Delta\bar{\omega}\end{bmatrix}.\label{38a}
\end{align} The inputs are the angles and magnitudes of the voltages on the $p$ buses of the study area, which are connected to the external area. The outputs are the rotor angle of the $p$ generators of the external area, which are connected to the study area. The step-wise procedure to generate equilibrium points and to reach equation (\ref{38a}) can be found in \cite{sauer1998power}-\nocite{anderson2008power,liu2009dynamic}\cite{chaniotis2001krylov}.\\
The nonlinear model (which is used for the damping controller design in this paper) is given by the following equations:
\begin{align}
\dot{\bar{\delta}}_i&=\bar{\omega}_i-\bar{\omega}_s,\label{39a}\\
\frac{2\bar{H}_i}{\bar{\omega}_s}\dot{\bar{\omega}}_i&=\bar{T}_i-(\bar{E}_{qi}^{\prime}-\bar{X}_{di}^{\prime} \bar{I}_{di})\bar{I}_{qi}-(\bar{E}_{di}^{\prime}+\bar{X}_{qi}^{\prime}\bar{I}_{qi})\bar{I}_{di},\label{40a}\\
\dot{\bar{E}}_{qi}^{\prime}&=-\frac{\bar{E}_{qi}^{\prime}}{\tau^{\prime}_{doi}}-\frac{(\bar{X}_{di}-\bar{X}_{di}^{\prime})\bar{I}_{di}}{\tau^{\prime}_{doi}}+\frac{\bar{E}_{fdi}}{\tau^{\prime}_{doi}},\label{41a}\\
\dot{\bar{E}}_{di}^{\prime}&=-\frac{\bar{E}^{\prime}_{di}}{\tau^{\prime}_{qoi}}-\frac{(\bar{X}_{qi}-\bar{X}_{qi}^{\prime})\bar{I}_{qi}}{\tau^{\prime}_{qoi}},\label{42a}\\
\dot{\bar{E}}_{fdi}&=-\frac{\big(\bar{K}_{Ei}+\bar{S}_{Ei}(\bar{E}_{fdi})\big)\bar{E}_{fdi}}{\bar{T}_{Ei}}+\frac{\bar{V}_{Ri}}{\bar{T}_{Di}},\label{43a}\\
\dot{\bar{V}}_{Ri}&=\frac{\bar{K}_{Ai}\bar{R}_{Fi}}{\bar{T}_{Ai}}+\frac{\bar{K}_{Ai}\bar{K}_{Fi}\bar{E}_{fdi}}{\bar{T}_{Fi}\bar{T}_{Ai}}-\frac{\bar{V}_{Ri}}{\bar{T}_{Ai}}\nonumber\\
&\hspace*{2.5cm}+\frac{\bar{K}_{Ai}(\bar{V}_{{ref}_i}-\bar{V}_i+\bar{V}_{s_i})}{\bar{T}_{A_i}},\label{44a}\\
\dot{\bar{R}}_{F_i}&=-\frac{\bar{R}_{F_i}}{\bar{T}_{F_i}}+\frac{\bar{K}_{F_i}\bar{E}_{fdi}}{\bar{T}_{F_i}^2}\label{45a}.
\end{align} Equations (\ref{39a})-(\ref{41a}) describe the dynamics of the machines where $\bar{E}_{di}^{\prime}$, $\bar{I}_{di}$, $\bar{X}_{di}$, and $\tau_{doi}^{\prime}$ are the emfs, currents, reactances, and time constants for the $d$-axis of the machine $i$; $\bar{E}_{qi}^{\prime}$, $\bar{I}_{qi}$, $\bar{X}_{qi}$, and $\tau_{qoi}^{\prime}$ are the emfs, currents, reactances, and time constants for the $q$-axis of the machine $i$; and $\bar{E}_{fdi}$ is the emf across the field winding of machine $i$. Equations (\ref{42a})-(\ref{45a}) describe the dynamics of the exciter, and the detailed description of the definitions of the parameters used in equations (\ref{42a})-(\ref{45a}) can be found in \cite{lee1992ieee}. The nonlinear equations (\ref{39a})-(\ref{45a}) can be linearized to obtain a $n^{th}$ order linear model, i.e.,
\begin{align}
\Delta\dot{x}=A\Delta x+B\Delta u, &&y=C\Delta x.\label{46a}
\end{align} where
\begin{align}
x=\begin{bmatrix}\Delta\bar{\delta}_i&\Delta\bar{\omega}_i&\Delta\bar{E}_{qi}^{\prime}&\Delta\bar{E}_{di}^{\prime}& \Delta\bar{E}_{fdi}&\Delta\bar{V}_{Ri}&\Delta\bar{R}_{F_i}\end{bmatrix}^T,\nonumber
\end{align}
$\Delta u=\begin{bmatrix}\Delta \bar{T}_i&\Delta\bar{V}_{{ref}_i}\end{bmatrix}^T$, and $\Delta y=\Delta\bar{\delta}_i$.\\
Let $G(s)$ be the $n^{th}$ order power system model with $m$ inputs and $p$ outputs, i.e.,
\begin{align}G(s)=C(sI-A)^{-1}B.\end{align}
The power system reduction problem is to find an $r^{th}$ ($r<<n$) order ROM $\tilde{G}(s)$ of the original model $G(s)$ such that the error $||G(s)-\tilde{G}(s)||$ is small in some defined sense. The projection based MOR techniques construct reduction subspaces $\tilde{V}$ and $\tilde{W}$, and the original system is projected onto that reduced subspace such that the dominant characteristics of the original system are retained in the ROM, i.e.,
\begin{align}
\tilde{G}(s)&=C\tilde{V}(sI-\tilde{W}^TA\tilde{V})^{-1}\tilde{W}^TB\nonumber\\
&=\tilde{C}(sI-\tilde{A})^{-1}\tilde{B}.
\end{align}
The important mathematical notations which are used throughout the text are tabulated in Table \ref{tab0}.
\begin{table}[!h]
\centering
\caption{Mathematical Notations}\label{tab0}
\begin{tabular}{|c|p{5cm}|}
\hline
Notation & Meaning \\ \hline
$\begin{bmatrix}\cdot\end{bmatrix}^*$   & Hermitian of the matrix.\\
$Re(\cdot)$   & Real part of the matrix.\\
   $\lambda_i(\cdot)$ & Eigenvalues of the matrix.\\
   $Ran(\cdot)$ & Range of the matrix.\\
   $\underset {i=1,\cdots,r}{span}\{\cdot\}$ & Span of the set of $r$ vectors.\\
   $||\cdot||_{\mathcal{H}_2}$  & $\mathcal{H}_2$-norm of the system.\\
   $||\cdot||_{\mathcal{H}_{2,\omega}}$ & Frequency-limited $\mathcal{H}_2$-norm of the system.\\
  $\mathcal{L}[\cdot]$ & Fr{\'e}chet derivative of the matrix logarithm.\\\hline
\end{tabular}
\end{table}
\subsection{Pseudo-Optimal Rational Krylov (PORK) Algorithm \cite{wolf2014h}}
Let $\sigma_i$ be the interpolation points in the tangential directions $\tilde{t}_i\in\mathbb{R}^{m\times1}$. Then $\tilde{G}(s)$ interpolates $G(s)$ at the interpolation points in the respective tangential directions, i.e., $\tilde{G}(\sigma_i)\tilde{t}_i=G(\sigma_i)\tilde{t}_i$ if the input rational Krylov subspace $\tilde{V}$ is given by
\begin{align}
Ran(\tilde{V})=\underset {i=1,\cdots,r}{span}\{(\sigma_iI-A)^{-1}B \tilde{t}_i\}.
\label{eq:3b}\end{align}
$\tilde{G}(s)$ interpolates $G(s)$ at the interpolation points in the respective tangential directions for any output rational Krylov subspace $\tilde{W}$ such that $\tilde{W}^T\tilde{V}=I$. Choose any $\tilde{W}$, for instance, $\tilde{W}=\tilde{V}$, and compute the following matrices
	\begin{align}
	\bar{E}&=\tilde{W}^T\tilde{V}, \hspace*{0.3cm}\bar{A}=\tilde{W}^TA\tilde{V},\hspace*{0.3cm}\bar{B}=\tilde{W}^TB,\\
	B_\bot&=B-\tilde{V}\bar{E}^{-1}\bar{B},\\
	\tilde{C}_t&=(B_\bot^TB_\bot)^{-1}B_\bot^T\big(A\tilde{V}-\tilde{V}\bar{E}^{-1}\bar{A}\big),\\
	S&=\bar{E}^{-1}\big(\bar{A}-\bar{B} \tilde{C}_t\big).
	\label{eq:7b}\end{align}
Then $\tilde{V}$ satisfies the following Sylvester equation:
\begin{align}
A\tilde{V}+\tilde{V}(-S)+B(-\tilde{C}_t)=0\nonumber
\end{align} where $\{\sigma_1,\cdots,\sigma_r\}$ are the eigenvalues of $S$. If the pair $(S,\tilde{C}_t)$ is observable, the ROM obtained with $\tilde{V}$ and $\tilde{W}$ can be parameterized in $\xi$ to obtain a family of ROMs which satisfy the interpolation condition $\tilde{G}(\sigma_i)\tilde{t}_i=G(\sigma_i)\tilde{t}_i$, i.e.,
\begin{align}
\tilde{A}=S+\xi\tilde{C}_t&& \tilde{B}=\xi&& \tilde{C}=CV.\nonumber
\end{align}
If $\xi$ is set to $\xi=-\tilde{Q}_t^{-1}\tilde{C}_t^T$ where $\tilde{Q}_t$ solves
\begin{align}
(-S^T)\tilde{Q}_t+\tilde{Q}_t(-S)+\tilde{C}_t^T\tilde{C}_t=0\nonumber,
\end{align} it ensures that $\tilde{G}(s)$ is a pseudo-optimal ROM for the problem $||G(s)-\tilde{G}(s)||^2_{\mathcal{H}_2}$ which satisfies the subset of first-order optimality conditions \cite{gugercin2008h_2}, i.e.,
\begin{align}
||G(s)-\tilde{G}(s)||^2_{\mathcal{H}_2}=||G(s)||^2_{\mathcal{H}_2}-||\tilde{G}(s)||^2_{\mathcal{H}_2}.\nonumber
\end{align}
\subsection{Frequency-Limited Balanced Truncation (FLBT) \cite{gawronski1990model}}
In \cite{gawronski1990model}, a frequency-limited generalization of balanced truncation \cite{moore1981principal} is presented which allows the user to specify the desired frequency region wherein superior accuracy is required. In FLBT \cite{gawronski1990model}, the standard controllability and observability Gramians, which are defined over the infinite frequency range, are replaced with the ones defined over the frequency region of interest. Let $P_\omega$ and $Q_\omega$ be the frequency-limited controllability and observability Gramains respectively defined over the desired frequency interval $[-\omega,\omega]$ rad/sec, i.e.,
\begin{align}
P_\omega&=\frac{1}{2\pi}\int_{-\omega}^{\omega}(j\nu I-A)^{-1}BB^T(j\nu I-A^T)^{-1}d\nu\nonumber\\
Q_\omega&=\frac{1}{2\pi}\int_{-\omega}^{\omega}(j\nu I-A^T)^{-1}C^TC(j\nu I-A)^{-1}d\nu\nonumber
\end{align} which solve the following Lyapunov equations:
\begin{align}
AP_\omega+P_\omega A^T+F(A)BB^T+BB^TF(A)^T&=0\nonumber\\
A^TQ_\omega+Q_\omega A+F(A)^TC^TC+C^TCF(A)&=0\nonumber
\end{align}
where
\begin{align}
F(A)&=\frac{1}{2\pi}\int_{-\omega}^{\omega}(j\nu I-A)^{-1}d\nu\\\nonumber
&=Re\Big(\frac{j}{\pi}ln(-j\omega I-A)\Big).\nonumber
\end{align}
The similarity transformation matrix $T_\omega$ is computed from $P_\omega$ and $Q_\omega$ as a contragradient transformation, i.e., $T_\omega^{-1}P_\omega T_\omega^{-T}=T_\omega^TQ_\omega T_\omega=diag( \bar{\sigma}_1, \bar{\sigma}_2,\cdots,\bar{\sigma}_n)$ where $\bar{\sigma}_1 \geq \bar{\sigma}_2 \geq \cdots \geq \bar{\sigma}_n$. The states associated with the least value of frequency-limited Hankel singular values $\bar{\sigma}_i$ are truncated. $\tilde{V}$ and $\tilde{W}$ in FLBT are computed as $\tilde{V}=T_{\omega}\tilde{Z}^T$ and $\tilde{W}=T_{\omega}^{-T}\tilde{Z}^T,$ respectively
where $\tilde{Z}=\begin{bmatrix}I_{r\times r} & 0_{r\times (n-r)}\nonumber\end{bmatrix}$.
\subsection{Frequency-Limited $\mathcal{H}_2$-Optimal MOR}
In \cite{petersson2014model}, a nonlinear optimization-based MOR algorithm is presented to achieve the local optimality for the problem $||G(s)-\tilde{G}(s)||^2_{\mathcal{H}_{2,\omega}}$ where
\begin{align}
||G(s)&-\tilde{G}(s)||^2_{\mathcal{H}_{2,\omega}}\nonumber\\
&=||G(s)||^2_{\mathcal{H}_{2,\omega}}+||\tilde{G}(s)||^2_{\mathcal{H}_{2,\omega}}-2\textnormal{ trace}(C\hat{P}_\omega \tilde{C}^T)\nonumber\\
&=||G(s)||^2_{\mathcal{H}_{2,\omega}}+||\tilde{G}(s)||^2_{\mathcal{H}_{2,\omega}}-2\textnormal{ trace}(\tilde{B}^T\hat{Q}_\omega B),\nonumber\\
||G(s)||^2_{\mathcal{H}_{2,\omega}}&=\frac{1}{2\pi}\int_{-\omega}^{\omega}\textnormal{trace}\big(G(j\nu)G(j\nu)^T\big)d\nu\nonumber\\
&=\frac{1}{2\pi}\int_{-\omega}^{\omega}\textnormal{trace}\big(G(j\nu)^TG(j\nu)\big)d\nu\nonumber\\
&=\textnormal{trace}(CP_\omega C^T)\nonumber\\
&=\textnormal{trace}(B^TQ_\omega B),\nonumber
\end{align}
and
\begin{align}
A\hat{P}_\omega+\hat{P}_\omega\tilde{A}^T+F(A)B\tilde{B}^T+B\tilde{B}^TF(\tilde{A})^T=0\nonumber\\
\tilde{A}^T\hat{Q}_\omega+\hat{Q}_\omega A+F(\tilde{A})^T\tilde{C}^TC+\tilde{C}^TCF(A)=0.\nonumber
\end{align}
$\tilde{G}(s)$ is a local optimum for this problem if the following first-order optimality conditions are satisfied:
\begin{align}
\hat{Q}_\omega B&=\tilde{Q}_\omega\tilde{B}\label{4a}\\
C\hat{P}_\omega&=\tilde{C}\tilde{P}_\omega\label{5a}\\
\hat{Q}^T_\omega\hat{P}_\omega+\tilde{Q}_\omega\tilde{P}_\omega&=Re\Big(\frac{j}{\pi}\mathcal{L}[-\tilde{A}-j\omega I,\mathcal{V}]\Big)^T\label{6a}
\end{align} where
\begin{align}
\tilde{A}\tilde{P}_\omega+\tilde{P}_\omega\tilde{A}+F(\tilde{A})\tilde{B}\tilde{B}^T+\tilde{B}\tilde{B}^TF(\tilde{A})^T=0,\label{eq:12b}\\
\tilde{A}^T\tilde{Q}_\omega+\tilde{Q}_\omega\tilde{A}+F(\tilde{A})^T\tilde{C}^T\tilde{C}+\tilde{C}^T\tilde{C}F(\tilde{A})=0,\\
\mathcal{V}=\tilde{C}^T\tilde{C}\tilde{P}_\omega-\tilde{C}^TC\hat{P}_\omega.
\end{align}
When $\tilde{G}(s)$ satisfies either (\ref{4a}) or (\ref{5a}), the following holds
\begin{align}
||G(s)&-\tilde{G}(s)||^2_{\mathcal{H}_{2,\omega}}=||G(s)||^2_{\mathcal{H}_{2,\omega}}-||\tilde{G}(s)||^2_{\mathcal{H}_{2,\omega}}.\nonumber
\end{align}
The nonlinear optimization algorithm to achieve a ROM which satisfies the above-mentioned first-order optimality conditions is computationally expensive and cannot be applied to large-scale systems. In \cite{vuillemin2014frequency}, the Gramian based optimality conditions (\ref{4a})-(\ref{6a}) are transformed into interpolation conditions; however, the iterative algorithm presented to achieve the local optimality requires the original system to be in pole-residue form which is only feasible for small-scale systems. Moreover, there is no guarantee on the convergence of the algorithm. In \cite{vuillemin2013h2}, a heuristic algorithm is presented, which produces a ROM with less $\mathcal{H}_{2,\omega}$-error, i.e., frequency-limited iterative rational Krylov algorithm (FLIRKA). Again, the convergence is not guaranteed in FLIRKA as well.
\section{Main Work}\label{Sec:III}
The analytical damping controller design procedures like LQG and $\mathcal{H}_\infty$ result in a controller whose order is greater than or equal to that of the power system model. To obtain a lower order controller, a ROM of the original model is first sought using MOR \cite{rogers2012power}-\nocite{pal2006robust}\cite{furuya2006robust}. It is stressed in \cite{sanchez1996power} that the ROM should preserve the critical modes and the frequency-domain behavior of the original system over the frequencies associated with the critical modes. These modes are generally poorly damped and can cause unstable operating conditions under heavy power transfers. Thus, the preservation of their identity in the ROM is critical for a good damping controller design that adds damping to these modes \cite{sanchez1996power}. Moreover, it is shown in \cite{scarciotti2015model,scarciotti2017low,yogarathinam2017new} that the preservation of these modes also improves the accuracy in the time-domain. We present a MOR algorithm for the power system reduction problem under consideration which not only preserves the specified modes of the original system, but it also ensures superior accuracy within the frequency region specified by the user. The proposed algorithm generates a ROM which satisfies a subset of the optimality conditions (\ref{4a})-(\ref{6a}). We call a ROM which satisfies either (\ref{4a}) or (\ref{5a}) as a frequency-limited pseudo-optimal ROM, and we name our algorithm as ``Frequency-limited Pseudo-optimal Rational Krylov algorithm (FLPORK)". FLPORK generates a ROM which has poles at the desired locations specified by the user.
\subsection{FLPORK}
We now present an algorithm that generates a frequency-limited pseudo-optimal ROM of $G(s)$. Let $\sigma_i$ be the interpolation points in the tangential directions $\tilde{t}_i\in\mathbb{R}^{m\times1}$. Define $B_\omega$, $G_\omega(s)$, $\tilde{G}_\omega(s)$, $S$, $\tilde{C}_t$, and $\hat{C}_t$ as
\begin{align}
B_\omega&=\begin{bmatrix}B &F(A)B\end{bmatrix},\hspace*{0.9cm} G_\omega(s)=C(sI-A)^{-1}B_\omega,\nonumber\\ \tilde{G}_\omega(s)&=C\tilde{V}(sI-\tilde{W}^*A\tilde{V})^{-1}\tilde{W}^*B_\omega=\tilde{C}(sI-\tilde{A})^{-1}\tilde{B}_\omega,\nonumber\\
S&=diag(\sigma_1,\cdots,\sigma_r),\hspace*{0.9cm}\tilde{C}_t=\begin{bmatrix}\tilde{t}_1&\cdots&\tilde{t}_r\end{bmatrix},\nonumber\\
\hat{C}_t&=\begin{bmatrix}\tilde{C}_tF(-S)\\\tilde{C}_t\end{bmatrix}=\begin{bmatrix}\hat{c}_1&\cdots&\hat{c}_r\end{bmatrix}.\label{eq:6}
\end{align}
$\tilde{G}_\omega(s)$ interpolates $G_\omega(s)$, i.e., $\tilde{G}_\omega(\sigma_i)\hat{c}_i=G_\omega(\sigma_i)\hat{c}_i$ for any $\tilde{W}$ such that $\tilde{W}^*\tilde{V}=I$ if
\begin{align}
\tilde{V}=\begin{bmatrix}(A-\sigma_1I)^{-1}B_\omega\hat{c}_1&\cdots&(A-\sigma_rI)^{-1}B_\omega\hat{c}_r\end{bmatrix}.
\end{align}
From the relation of Krylov subspaces and Sylvester equations \cite{panzer2014model}, it can be noted that $\tilde{V}$ solves the following Sylvester equation:
\begin{align}
A\tilde{V}+\tilde{V}(-S)+B_\omega(-\hat{C}_t)=0.\label{eq:8}
\end{align}
If all the eigenvalues of $S$ have a positive real part and the pair $(S,\hat{C}_t)$ is observable, $(\tilde{A},\tilde{B}_\omega,\tilde{C})$ obtained with $\tilde{V}$ and $\tilde{W}$ can be parameterized in $\xi$ to obtain a family of ROMs \cite{ahmad2011krylov}, i.e.,
\begin{align}
\tilde{A}=S+\xi\hat{C}_t, && \tilde{B}_\omega=\xi, && \tilde{C}=C\tilde{V}
\end{align} which satisfy the interpolation condition $\tilde{G}_\omega(\sigma_i)\hat{c}_i=G_\omega(\sigma_i)\hat{c}_i$.
This can be readily verified by multiplying (\ref{eq:8}) with $\tilde{W}^*$ from the left; see \cite{panzer2014model}. Set $\xi$ to
\begin{align}
\xi=\begin{bmatrix}-(\tilde{Q}_{s\omega})^{-1}\tilde{C}_t^*&-(\tilde{Q}_{s\omega})^{-1}F^*(-S)\tilde{C}_t^*\end{bmatrix}
\end{align} where
\begin{align}
(-S^*)\tilde{Q}_{s\omega}+\tilde{Q}_{s\omega}& (-S)+\nonumber\\
&F^*(-S)\tilde{C}_t^*\tilde{C}_t+\tilde{C}_t^*\tilde{C}_tF(-S)=0.\label{eq:11}
\end{align}
Then, ROM $(\tilde{A},\tilde{B},\tilde{C})$ in FLPORK is obtained by removing $-(\tilde{Q}_{s\omega})^{-1}F^*(-S)\tilde{C}_t^*$ from $\tilde{B}_\omega$, i.e.,
\begin{align}
\tilde{A}&=S-(\tilde{Q}_{s\omega})^{-1}\tilde{C}_t^*\tilde{C}_tF(-S)-(\tilde{Q}_{s\omega})^{-1}F^*(-S)\tilde{C}_t^*\tilde{C}_t,\nonumber\\
\tilde{B}&=-(\tilde{Q}_{s\omega})^{-1}\tilde{C}_t^*,\hspace*{0.5cm}\tilde{C}=CV.\label{eq:12}
\end{align}
\textbf{Theorem 1:} If $(\tilde{A},\tilde{B},\tilde{C})$ is defined as in equation (\ref{eq:12}), then $\tilde{G}(s)$ has the following properties:\\
(i) $\tilde{G}(s)$ has poles at the mirror images of the interpolation points $\sigma_i$.\\
(ii) $(\tilde{Q}_{s\omega})^{-1}$ is the frequency-limited controllability Gramian of the pair $(\tilde{A},\tilde{B})$.\\
(iii) $C\hat{P}_\omega=\tilde{C}\tilde{P}_\omega$.\\
(iv) $\tilde{t}_i$ is the input-residual of $\tilde{G}(s)$.\\
\textbf{Proof:} (i) By multiplying $(\tilde{Q}_{s\omega})^{-1}$ from the left side of equation (\ref{eq:11}) yields
\begin{align}
-(\tilde{Q}_{s\omega})^{-1}S^*\tilde{Q}_{s\omega}-S+(\tilde{Q}_{s\omega})^{-1}&F^*(-S)\tilde{C}_t^*\tilde{C}_t\nonumber\\
&+(\tilde{Q}_{s\omega})^{-1}\tilde{C}_t^*\tilde{C}_tF(-S)=0\nonumber\\
-(\tilde{Q}_{s\omega})^{-1}S^*\tilde{Q}_{s\omega}-\tilde{A}&=0.\nonumber
\end{align}
Thus $\tilde{A}=-(\tilde{Q}_{s\omega)}^{-1}S^*\tilde{Q}_{s\omega}$ and hence $\lambda_i(\tilde{A})=-\lambda_i(S^*)$.\\
(ii) The frequency-limited controllability Gramian $\tilde{P}_\omega$ of the pair $(\tilde{A},\tilde{B})$ solves equation (\ref{eq:12b}). By pre- and post-multiplying equation (\ref{eq:12b}) with $\tilde{Q}_{s\omega}$, by putting $\tilde{A}=-(\tilde{Q}_{s\omega})^{-1}S^*\tilde{Q}_{s\omega}$ and $\tilde{B}=-(\tilde{Q}_{s\omega})^{-1}\tilde{C}_t^*$, and also by noting that $\tilde{Q}_{s\omega} F(\tilde{A})(\tilde{Q}_{s\omega})^{-1}=F^*(-S)$, equation (\ref{eq:12b}) becomes
\begin{align}
(-S^*)\tilde{Q}_{s\omega}\tilde{P}_\omega\tilde{Q}_{s\omega}&+\tilde{Q}_{s\omega}\tilde{P}_\omega\tilde{Q}_{s\omega} (-S)\nonumber\\
&+F^*(-S)\tilde{C}_t^*\tilde{C}_t+\tilde{C}_t^*\tilde{C}_tF(-S)=0.\nonumber
\end{align}
Due to uniqueness, $\tilde{Q}_{s\omega}\tilde{P}_\omega\tilde{Q}_{s\omega}=\tilde{Q}_{s\omega}$, $\tilde{Q}_{s\omega}\tilde{P}_\omega=I$, and $\tilde{P}_\omega=(\tilde{Q}_{s\omega})^{-1}$.\\
(iii) Consider the following equation:
\begin{align}
&A\tilde{V}\tilde{P}_\omega+\tilde{V}\tilde{P}_\omega\tilde{A}^*+F(A)B\tilde{B}^*+B\tilde{B}^*F^*(\tilde{A})\nonumber\\
&=[\tilde{V}S+B_\omega\hat{C}_t]\tilde{P}_\omega+\tilde{V}\tilde{P}_\omega[-(\tilde{Q}_{s\omega})^{-1}S^*\tilde{Q}_{s\omega}]^*\nonumber\\
&\hspace*{3.5cm}-F(A)B\tilde{C}_t\tilde{P}_\omega-B\tilde{C}_tF(-S)\tilde{P}_\omega\nonumber\\
&=\tilde{V}S\tilde{P}_\omega+F(A)B\tilde{C}_t\tilde{P}_\omega+B\tilde{C}_tF(-S)\tilde{P}_\omega-\tilde{V}S\tilde{P}_\omega\nonumber\\
&\hspace*{3.5cm}-F(A)B\tilde{C}_t\tilde{P}_\omega-B\tilde{C}_tF(-S)\tilde{P}_\omega\nonumber\\
&=0.\nonumber
\end{align}
Due to uniqueness, $\tilde{V}\tilde{P}_\omega=\hat{P}_\omega$, and thus, $C\hat{P}_\omega=\tilde{C}\tilde{P}_\omega$.\\
(iv) $\tilde{A}=(-\tilde{Q}_{s\omega})^{-1}(-S^*)(-\tilde{Q}_{s\omega})$ is actually the spectral factorization of $\tilde{A}$. Moveover, $\tilde{B}=(-\tilde{Q}_{s\omega})^{-1}\begin{bmatrix}\tilde{t}_1&\cdots&\tilde{t}_r\end{bmatrix}^T$. Thus, $\tilde{t}_i$ is the input-residual of $\tilde{G}(s)$.\\
\\A dual result also exists wherein $\tilde{W}$ is fixed with an arbitrary choice of $\tilde{V}$, and then the ROM is parameterized to achieve frequency-limited pseudo-optimality. We call it Output-FLPORK (O-FLPORK) to differentiate with the previous case. There is some abuse in the mathematical notations, but the context leaves no ambiguity. Let $\sigma_i$ be the interpolation points in the tangential directions $\hat{t}_i\in\mathbb{R}^{1\times p}$. Define $C_\omega$, $\bar{G}_\omega(s)$, $\tilde{\bar{G}}_\omega(s)$, $S$, $\tilde{B}_t$, and $\hat{B}_t$ as
\begin{align}
C_\omega&=\begin{bmatrix}C \\ CF(A)\end{bmatrix},\hspace*{0.9cm} \bar{G}_\omega(s)=C_\omega(sI-A)^{-1}B,\nonumber\\ \tilde{\bar{G}}_\omega(s)&=C_\omega\tilde{V}(sI-\tilde{W}^*A\tilde{V})^{-1}\tilde{W}^*B=\tilde{C}_\omega(sI-\tilde{A})^{-1}\tilde{B},\nonumber\\
S&=diag(\sigma_1,\cdots,\sigma_r)\hspace*{0.5cm}\tilde{B}_t=\begin{bmatrix}\hat{t}_1^*&\cdots&\hat{t}_r^*\end{bmatrix}^T \nonumber\\
\hat{B}_t&=\begin{bmatrix}F(-S)\tilde{B}_t&\tilde{B}_t\end{bmatrix}=\begin{bmatrix}\hat{b}_1^*&\cdots&\hat{b}_r^*\end{bmatrix}^T.
\label{eq:14}\end{align}
$\tilde{\bar{G}}_\omega(s)$ interpolates $\bar{G}_\omega(s)$, i.e., $\hat{b}_i\tilde{\bar{G}}_\omega(\sigma_i)=\hat{b}_i\bar{G}_\omega(\sigma_i)$ for any $\tilde{V}$ such that $\tilde{W}^*\tilde{V}=I$ if
\begin{align}
\tilde{W}=\big[(A-\sigma_1I)^{-*}C_\omega^*\hat{b}_1^*\cdots(A-\sigma_rI)^{-*}C_\omega^*\hat{b}_r^*\big].
\end{align}
From the relation of Krylov subspaces and Sylvester equations \cite{panzer2014model}, it can be noted that $\tilde{W}$ solves the following Sylvester equation:
\begin{align}
\tilde{W}^*A+(-S)\tilde{W}^*+(-\hat{B}_t)(C_\omega)=0.\label{eq:16}
\end{align}
If all the eigenvalues of $S$ have positive real part and the pair $(S,\hat{B}_t)$ is controllable, $(\tilde{A},\tilde{B},\tilde{C}_\omega)$ obtained with $\tilde{V}$ and $\tilde{W}$ can be parameterized in $\xi$ to obtain a family of ROMs which satisfy the interpolation condition $\hat{b}_i\tilde{\bar{G}}_\omega(\sigma_i)=\hat{b}_i\bar{G}_\omega(\sigma_i)$ \cite{ahmad2011krylov}, i.e.,
\begin{align}
\tilde{A}=S+\hat{B}_t\xi, && \tilde{B}=\tilde{W}^*B, && \tilde{C}_\omega=\xi.
\end{align}
This can be readily verified by multiplying (\ref{eq:16}) with $\tilde{V}$ from the right. Set $\xi$ to
\begin{align}
\xi=\begin{bmatrix}-\tilde{B}_t^*(\tilde{P}_{s\omega})^{-1}\\ -\tilde{B}_t^*F^*(-S)(\tilde{P}_{s\omega})^{-1}\end{bmatrix}
\end{align} where
\begin{align}
(-S)\tilde{P}_{s\omega}+\tilde{P}_{s\omega} (-S^*)+F(-S)\tilde{B}_t\tilde{B}_t^*+\tilde{B}_t\tilde{B}_t^*F^*(-S)=0.\label{eq:19}
\end{align}
Then, ROM $(\tilde{A},\tilde{B},\tilde{C})$ in O-FLPORK is obtained by removing $-\tilde{B}_t^*F^*(-S)(\tilde{P}_{s\omega})^{-1}$ from $\tilde{C}_\omega$, i.e.,
\begin{align}
\tilde{A}&=S-F(-S)\tilde{B}_t\tilde{B}_t^*\tilde{P}_{s\omega}^{-1}-\tilde{B}_t\tilde{B}_t^*F^*(-S)\tilde{P}_{s\omega}^{-1},\nonumber\\
\tilde{B}&=\tilde{W}^*B,\hspace*{0.5cm}\tilde{C}=-\tilde{B}_t^*\tilde{P}_{s\omega}^{-1}.\label{eq:20}
\end{align}
\textbf{Theorem 2:} If $(\tilde{A},\tilde{B},\tilde{C})$ is defined as in equation (\ref{eq:20}), then $\tilde{G}(s)$ has the following properties:\\
(i) $\tilde{G}(s)$ has poles at the mirror images of the interpolation points $\sigma_i$.\\
(ii) $(\tilde{P}_{s\omega})^{-1}$ is the frequency-limited observability Gramian of the pair $(\tilde{A},\tilde{C})$.\\
(iii) $\hat{Q}_\omega B=\tilde{Q}_\omega\tilde{B}$.\\
(iv) $\hat{t}_i$ is the output-residual of $\tilde{G}(s)$.\\
\textbf{Proof:} The proof is similar to that of Theorem 1 and hence omitted for brevity.\\
\\\textit{Remark 1:} If the interpolation points are selected as the mirror images of $r$ specified electromechanical poles of $G(s)$, the ROM preserves these modes. Moreover, if $\omega$ is selected such that $[-\omega,\omega]$ contains the frequency region wherein the local and interarea oscillations lie, the ROM accurately captures these oscillations in its frequency-domain response.\\
\textit{Remark 2:} For simplicity, it is assumed throughout the paper that the desired frequency interval is $[-\omega,\omega]$. However, it can be any symmetric interval, i.e., $[-\omega_2,-\omega_1]\cup[\omega_1,\omega_2]$ rad/sec where $\omega_2>\omega_1>0$. In that case, $F(A)$ and $F(-S)$ become
\begin{align}
F(A)&=\frac{1}{2\pi}\Big(\int_{-\omega_2}^{\omega_2}(j\nu I-A)^{-1}d\nu-\int_{-\omega_1}^{\omega_1}(j\nu I-A)^{-1}d\nu\Big)\nonumber\\
F(-S)&=\frac{1}{2\pi}\Big(\int_{-\omega_2}^{\omega_2}(j\nu I+S)^{-1}d\nu-\int_{-\omega_1}^{\omega_1}(j\nu I+S)^{-1}d\nu\Big).\nonumber
\end{align}
\subsection{Connection with FLIRKA \cite{vuillemin2013h2}}
Let $\tilde{G}(s)$ be the ROM obtained after FLIRKA converges, and it is represented by the following pole-residue form
\begin{align}
\tilde{G}(s)=\sum_{i=1}^{r}\frac{\bar{l}_i\bar{r}_i^*}{s-\lambda_i}.\nonumber
\end{align}
Then the reduction subspaces generated by FLIRKA in the last iteration ensure that the following interpolatory conditions are satisfied:
\begin{align}
G_\omega(-\lambda_i)\hat{r}_i&=\tilde{G}_\omega(\lambda_i)\hat{r}_i\label{24a}\\
\hat{l}_i^*\bar{G}_\omega(-\lambda_i)&=\hat{l}_i^*\tilde{\bar{G}}_\omega(\lambda_i)\label{25a}
\end{align} where
\begin{align}
\begin{bmatrix}\hat{r}_1&\cdots&\hat{r}_r\end{bmatrix}&=\begin{bmatrix}\begin{bmatrix}\bar{r}_i&\cdots&\bar{r}_r\end{bmatrix}\times F(D)\\\begin{bmatrix}\bar{r}_i&\cdots&\bar{r}_r\end{bmatrix}\end{bmatrix}\nonumber\\
\begin{bmatrix}\hat{l}_1\\\vdots\\\hat{l}_r\end{bmatrix}&=\begin{bmatrix} F(D)\times\begin{bmatrix}\bar{l}_i\\\vdots\\\bar{l}_r\end{bmatrix}&\begin{bmatrix}\bar{l}_i\\\vdots\\\bar{l}_r\end{bmatrix}\end{bmatrix}\nonumber\\
D&=diag(\lambda_1,\cdots,\lambda_r).\nonumber\end{align}
In other words, $\tilde{G}(s)$ is an input- and output- frequency-limited pseudo optimal ROM. This explains the reason why FLIRKA ensures good $\mathcal{H}_{2,\omega}$ accuracy if the convergence is achieved. Note that FLIRKA is proposed heuristically based on experimental results. FLPORK and O-FLPORK can thus be seen as iteration-free algorithms which judiciously place the poles of $\tilde{G}(s)$ at the mirror images of the interpolation points make the tangential directions as input and output residuals respectively. Therefore, FLPORK and O-FLPORK satisfy (\ref{24a}) and (\ref{25a}), respectively.
\subsection{Choice of Modes to be Preserved}
There is no theoretical guarantee that preserving a particular mode of the original system will surely ensure less error. However, some general and practical guidelines regarding the preservation of the ``right" set of modes of the original system are reported in the literature. For instance, it is suggested in \cite{sanchez1996power} to preserve the local and interarea modes in the ROM for accurately capturing the local and interarea oscillations. This is particularly important when the ROM is used for obtaining a damping controller to reduce the local and interarea oscillations. It is argued in \cite{scarciotti2015model,scarciotti2017low} that a good time-domain accuracy can be achieved if the slowest and most poorly damped modes of the original model are preserved in the ROM. The lightly damped modes in the frequency range $[0,2]$ Hz are called electromechanical modes \cite{tp4622012identification} and can easily be captured using Subspace Accelerated Rayleigh Quotient Iteration (SARQI) algorithm \cite{rommes2010computing}. The peaks and dips in the frequency response of a system are associated with the modes with large residuals which can be easily captured using Subspace Accelerated MIMO Dominant Pole Algorithm (SAMDP) \cite{rommes2006efficient}. It is shown in \cite{rommes2006efficient} that the preservation of these modes in the ROM yields overall good accuracy over the entire frequency range. These guidelines should be followed when choosing the interpolation points in FLPORK and O-FLPORK to obtain a high-fidelity ROM.
\subsection{Algorithmic Aspects}
We allowed the state-space matrices to be complex in the previous subsection; however, one can obtain a real ROM for a real original model. $F(A)$ is a real matrix when the desired frequency region is symmetric \cite{petersson2014model,benner2016frequency} $\big([-\omega_2,-\omega_1]\cup[\omega_1,\omega_2]\big)$, i.e.,
\begin{align}
F(A)=Re\Big(\frac{j}{\pi}ln\big((j\omega_1I+A)^{-1}(j\omega_2I+A)\big)\Big).\nonumber
\end{align}
This leads to a real $B_\omega$. The real $S$ and $\tilde{C}_t$ can be obtained from equations (\ref{eq:3b})-(\ref{eq:7b}) which lead to the real $\hat{C}_t$, $\tilde{Q}_{s\omega}$, and $\tilde{V}$. Finally, the ROM is obtained as
\begin{align}
\tilde{A}&=-\tilde{Q}_{s\omega}^{-1}S^T\tilde{Q}_{s\omega},\hspace{1cm}\tilde{B}=-\tilde{Q}_{s\omega}^{-1}\tilde{C}_t^T,\nonumber\\
\tilde{C}&=C\tilde{V}.\nonumber
\end{align}
Similarly, $C_\omega$ is a real matrix when the desired frequency region is symmetric, i.e., $[-\omega_2,-\omega_1]\cup[\omega_1,\omega_2]$. The real $S$ and $\tilde{B}_t$ which have the information of the interpolation points $\sigma_i$ and tangential directions $\hat{t}_i$ encoded in them can be computed by the following steps. Compute $\bar{W}$ as
\begin{align}
Ran(\bar{W})=\underset {i=1,\cdots,r}{span}\{(\sigma_iI-A^T)^{-1}C^T\hat{t}_i^T\}.\nonumber
\end{align}
Choose any $\bar{V}$, for instance, $\bar{V}=\bar{W}$. Then compute the following matrices
\begin{align}
\bar{E}&=\bar{W}^T\bar{V}, \hspace*{0.3cm}\bar{A}=\bar{W}^TA\bar{V},\hspace*{0.3cm}\bar{C}=C\bar{V},\nonumber\\
C_\bot&=C-\bar{C}\bar{E}^{-1}\bar{W}^T\nonumber\\
\tilde{B}_t&=\big(\bar{W}^TA-\bar{A}\bar{E}^{-1}\bar{W}^T\big)C_\bot^T(C_\bot C_\bot^T)^{-1}\nonumber\\
S&=\big(\bar{A}-\tilde{B}_t\bar{C}\big)\bar{E}^{-1}.\nonumber
\end{align}
The real $S$ and $\tilde{B}_t$ lead to the real $\hat{B}_t$, $\tilde{P}_{s\omega}$, and $\tilde{W}$. Then the ROM is obtained as
\begin{align}
\tilde{A}&=-\tilde{P}_{s\omega}S^T\tilde{P}_{s\omega}^{-1}, \hspace{1cm}\tilde{B}=\tilde{W}^TB,\nonumber\\ \tilde{C}&=-\tilde{B}_t^T\tilde{P}_{s\omega}^{-1}.\nonumber
\end{align}
\subsection{Computational Aspects}
FLBT \cite{gawronski1990model} becomes computationally expensive in a large-scale setting because it requires the solution of two large-scale Lyapunov equations. In \cite{benner2016frequency}, FLBT is generalized using a low-rank approximation of Lyapunov equations to extend its applicability to large-scale systems. FLPORK and O-FLPORK do not involve large-scale Lyapunov equations like FLBT \cite{gawronski1990model}. However, as argued in \cite{benner2016frequency}, $F(A)$ may become computationally expensive in a large-scale setting. In \cite{benner2016frequency}, various computationally efficient approaches to compute $F(A)$ are discussed which also includes quadrature rules for numerical integration. The computational cost in these methods directly depends on the number of quadrature nodes. A trade-off can be done between the accuracy and computational cost depending on the size of $A$ in a particular problem, and an appropriate number of quadrature nodes can be selected to compute the integral $F(A)$ within the admissible time. Once $F(A)$ is computed, the main computational effort in FLPORK and O-FLPORK is spent on the solution of ``\textit{sparse-dense}" Sylvester equations (\ref{eq:8}) and (\ref{eq:16}) respectively because the Lyapunov equations (\ref{eq:11}) and (\ref{eq:19}) are small-scale equations. Equations (\ref{eq:8}) and (\ref{eq:16}) have large but sparse matrices $A$, $B_\omega$, and $C_\omega$ owing to the sparse structure of power system state-space model \cite{freitas2008gramian}, and dense but small matrices $S$, $\hat{C}_t$ and $\hat{B}_t$. As shown in \cite{wolf2014h}, the solution of ``\textit{sparse-dense}" Sylvester equations can be obtained within admissible time as long as $r<<n$ which is the situation in MOR and therefore, the Krylov subspaces $\tilde{V}$ and $\tilde{W}$ can be computed easily by either using direct or iterative methods. Thus, FLPORK and O-FLPORK are easily applicable to large-scale power systems.
\section{Applications}
In this section, we demonstrate the applications of FLPORK on three interconnected power system models. The first model is an interconnection of the IEEE $145$-bus $50$-machine system with the New England Test System-New York Power System (NETS-NYPS) $68$-bus $16$-machine system. The second model is the interconnection of the Northeastern Power Coordinating Council (NPCC) $140$-bus $48$-machine system with the IEEE $145$-bus $50$-machine system. The third model is the IEEE $145$-bus $50$-machine system. We first compare the performance of FLPORK with SARQI \cite{rommes2010computing} -based modal truncation, PORK \cite{wolf2014h}, FLIRKA \cite{vuillemin2013h2}, FLBT \cite{gawronski1990model}, and finite-frequency modal iterative rational Krylov algorithm (FFMIRKA) \cite{zulfiqar2019finite} both in the frequency and time domains on two power system reduction problems. Next, we design a reduced-order $\mathcal{H}_\infty$ damping controller for the IEEE $145$-bus $50$-machine system using FLPORK and compare its performance with SARQI \cite{rommes2010computing} -based modal truncation, PORK \cite{wolf2014h}, FLIRKA \cite{vuillemin2013h2}, FLBT \cite{gawronski1990model}, and FFMIRKA \cite{zulfiqar2019finite}. The results of O-FLPORK are indistinguishable from FLPORK, and hence, only results of FLPORK are shown. All the experiments are performed on a laptop with Intel Core M-5Y10c processor, 8GB of RAM, and Windows 8 operating system.
\subsection{Power System Reduction}
\begin{figure}[!h]
\centering
\includegraphics[width=8.2cm]{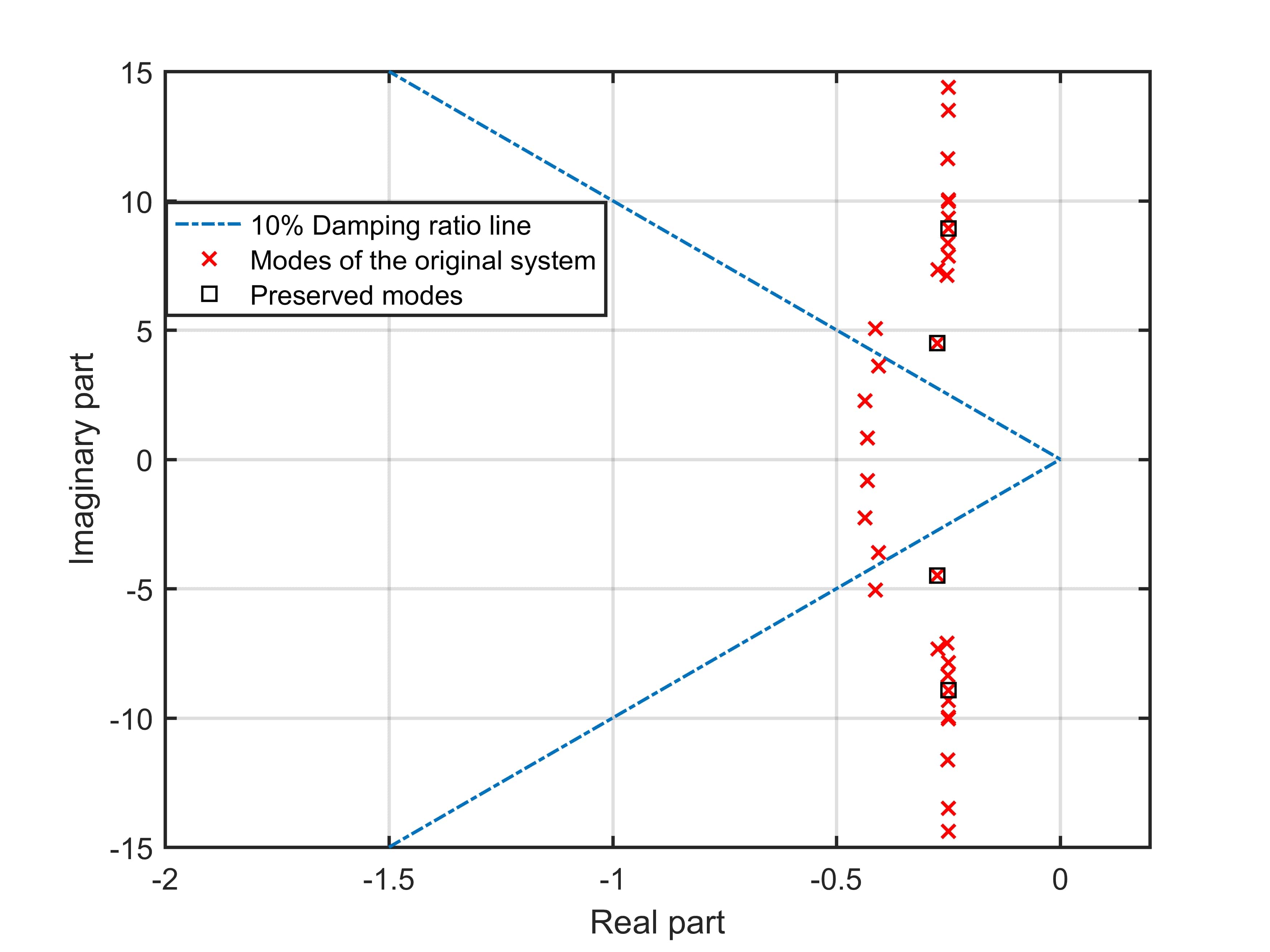}
\caption{The rightmost modes of NETS-NYPS}\label{fig:2}
\end{figure}

\begin{figure}[!h]
\centering
\includegraphics[width=8.4cm]{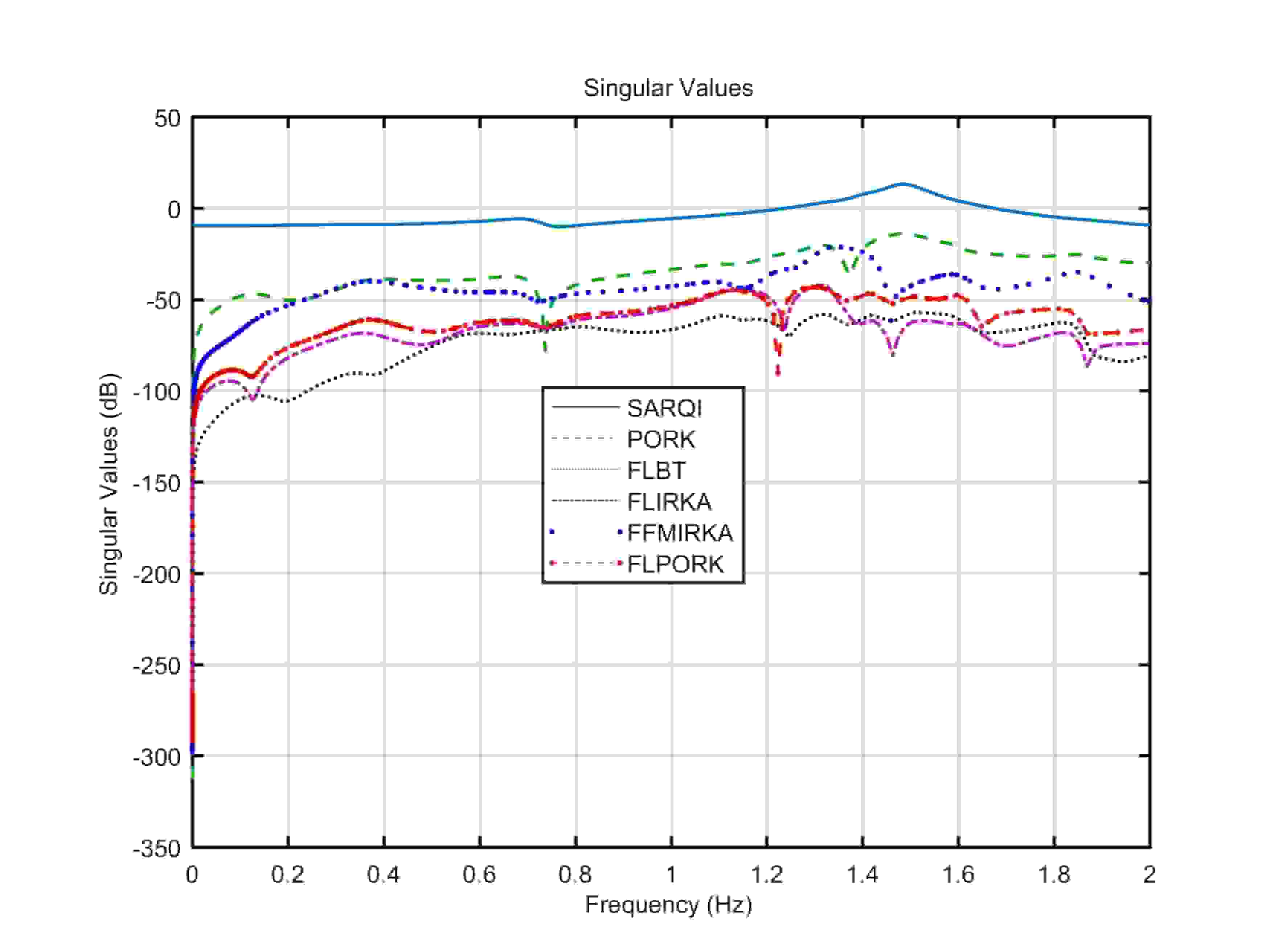}
\caption{Frequency domain error within $[0,2]$ Hz}\label{fig:3}
\end{figure}

\begin{figure}[!h]
\centering
\includegraphics[width=8.4cm]{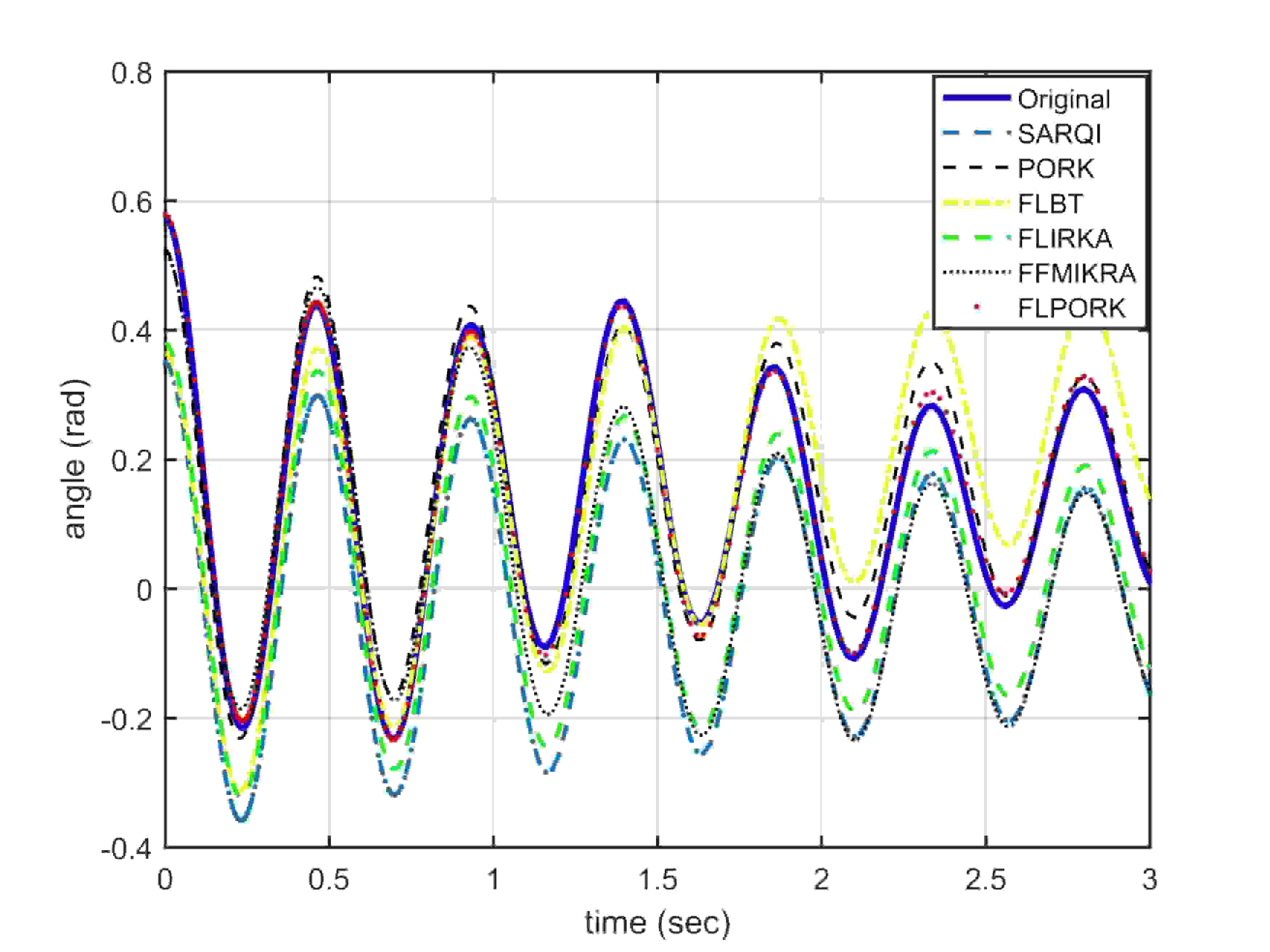}
\caption{Rotor angle of generator-$1$ of NETS-NYPS test system}\label{fig:4}
\end{figure}

\begin{figure}[!h]
\centering
\includegraphics[width=8.2cm]{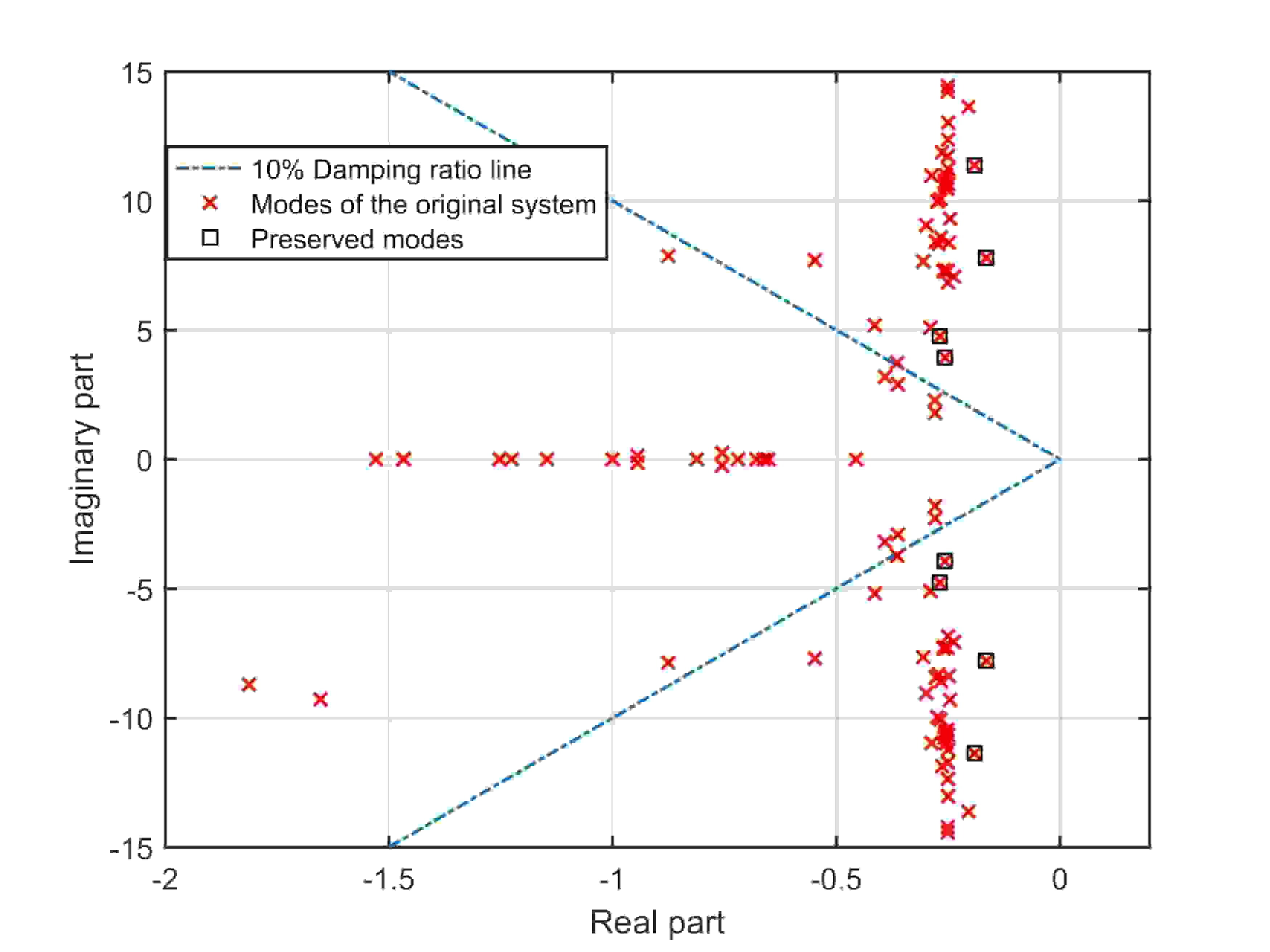}
\caption{The rightmost modes of IEEE $50$-machine test system}\label{fig:5a}
\end{figure}

\begin{figure}[!h]
\centering
\includegraphics[width=8.4cm]{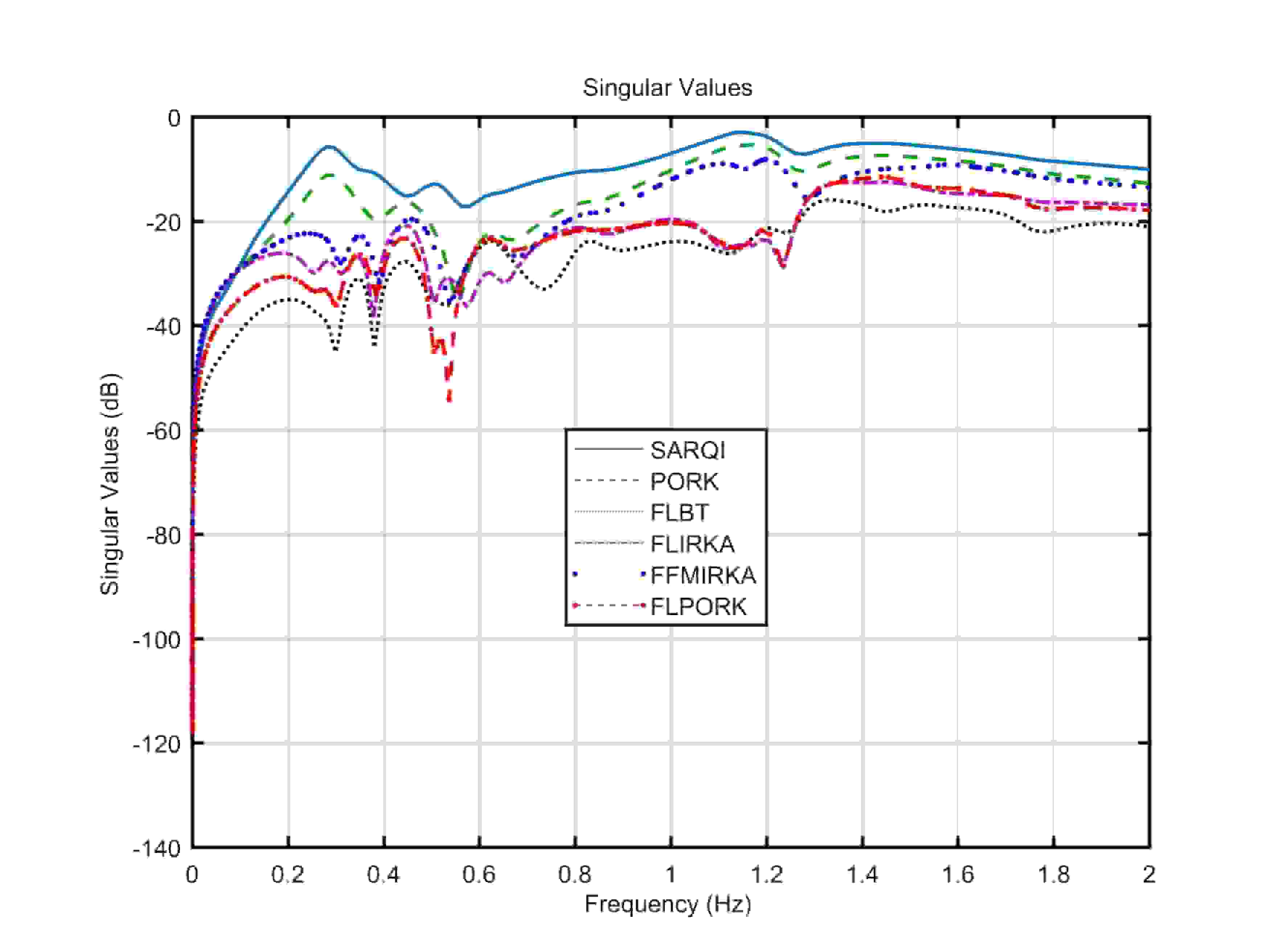}
\caption{Frequency domain error within $[0,2]$ Hz}\label{fig:6a}
\end{figure}

\begin{figure}[!h]
\centering
\includegraphics[width=8.4cm]{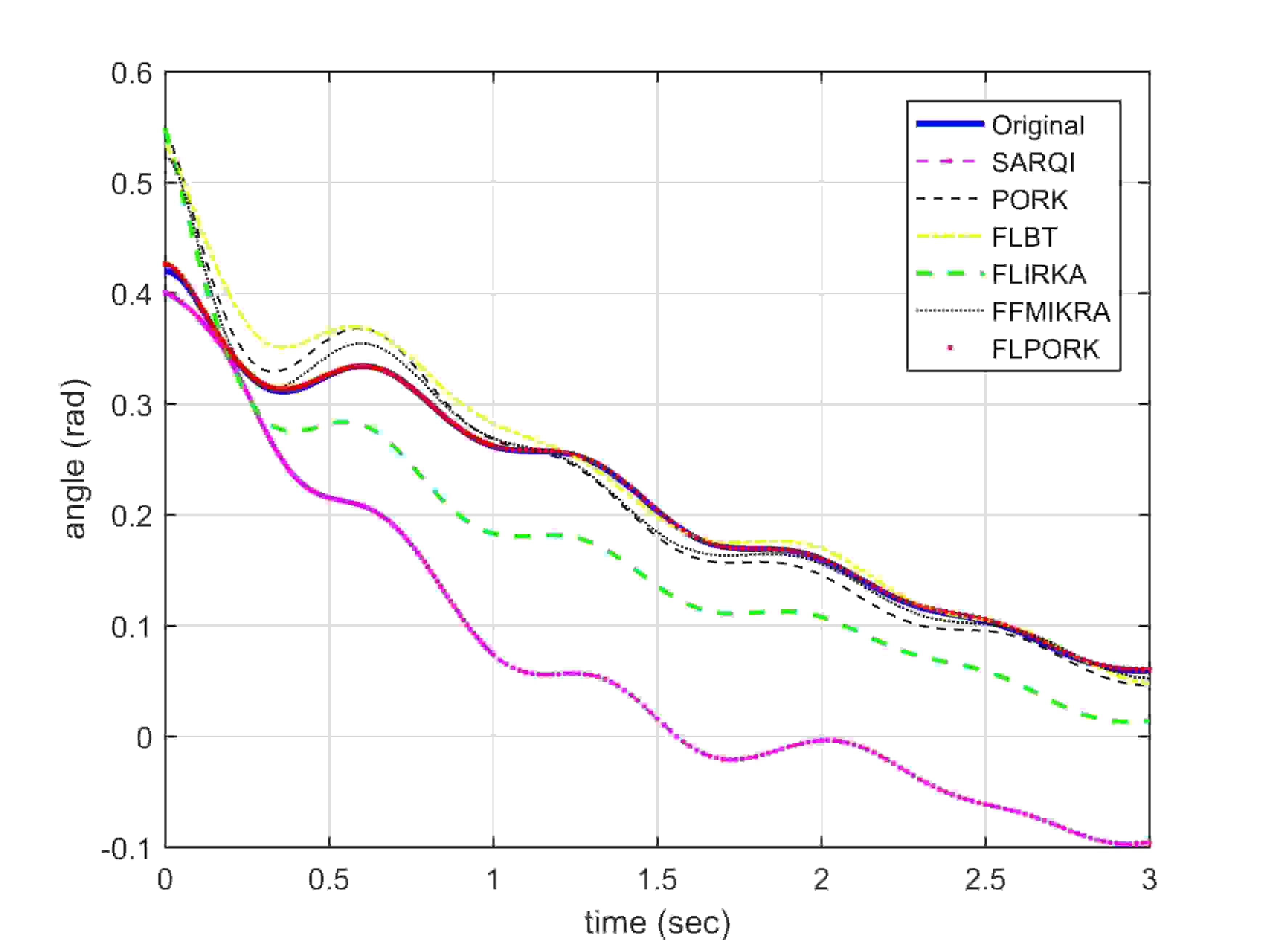}
\caption{Rotor angle of generator-$1$ of IEEE $145$-bus $50$-machine test system}\label{fig:7a}
\end{figure}

\begin{figure}[!h]
\centering
\includegraphics[width=8.4cm]{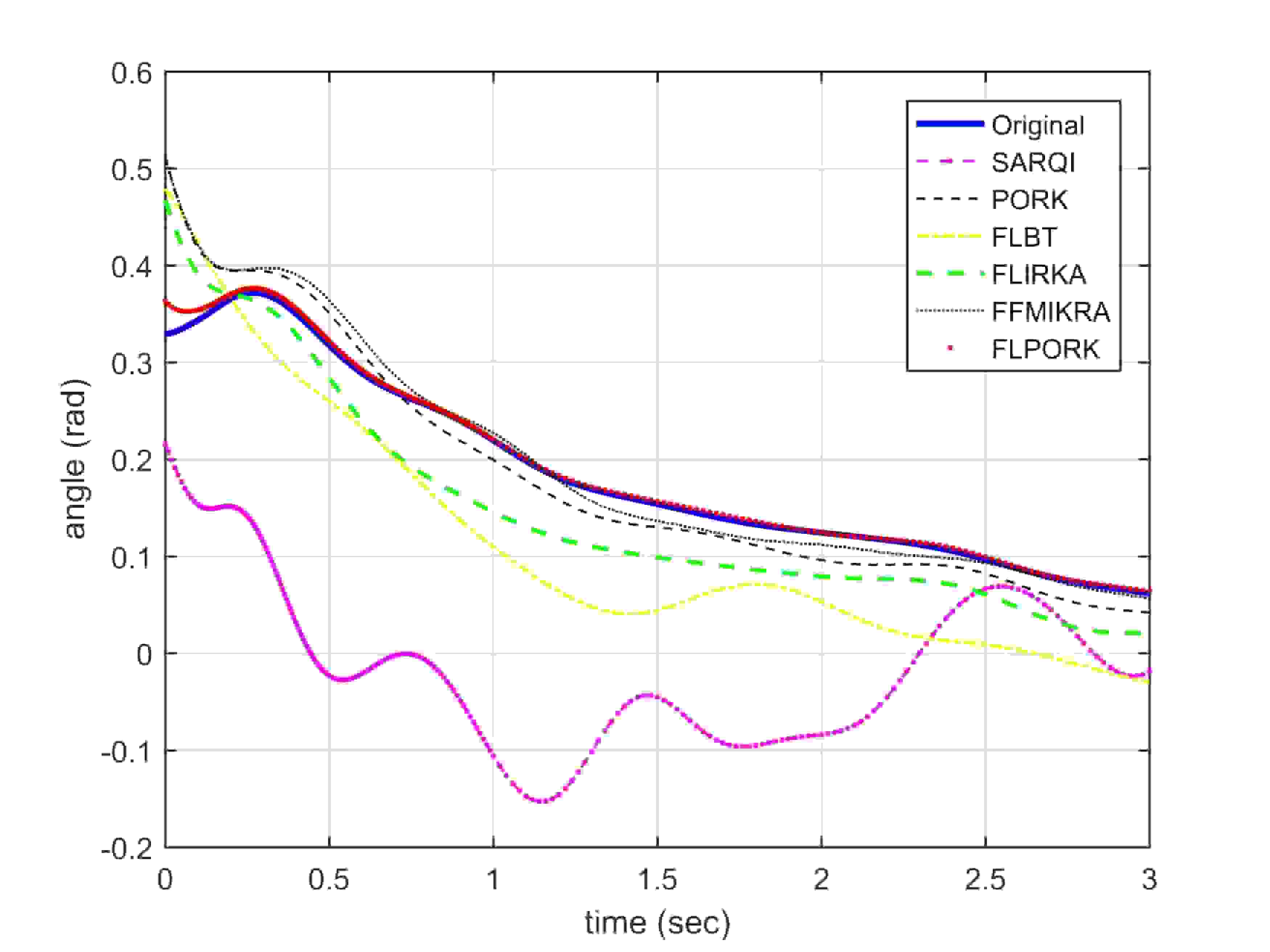}
\caption{Rotor angle of generator-$2$ of IEEE $145$-bus $50$-machine test system}\label{fig:8a}
\end{figure}

\textit{\textbf{Test 1: NETS-NYPS connected to IEEE 145-bus 50-machine system via One Tie Line:}} In this experiment, the NETS-NYPS $16$-machine, $68$-bus system taken from \cite{chow1992toolbox} is considered as the external area described by the $32^{nd}$ order linear model in equation (\ref{38a}). The study area is the IEEE $145$-bus, $50$-machine system also taken from \cite{chow1992toolbox}. Bus-$53$ of the external area is connected to the bus-$60$ of the study area via one tie line. The inputs are the voltage magnitude and angle on bus-$60$ of the study area, and the output is the rotor angle of generator-$1$ which is connected to bus-$53$ of the external area. The study area retains its nonlinear description. The rightmost most poles of the external area are plotted in Fig. \ref{fig:2}, and it can be seen that it has several poorly damped modes in its model.
 A $10$th order ROM of the external area is generated by SARQI -based modal truncation, PORK, FLIRKA, FLBT, FFMIRKA, and FLPORK. SARQI, PORK, FFMIRKA, and FLPORK preserve the most poorly damped two interarea and two local modes of the original system, i.e., $-0.2748\pm j4.4888$ and $-0.25\pm j8.9339$, respectively to capture the interarea and local oscillations in the ROM. SARQI additionally preserves the following modes: $-0.25\pm j14.3865$, $-0.25\pm j13.4964$, and $-0.2514\pm j11.6271$ to make the ROM a $10^{th}$ order model. The mirror images of the poorly damped modes may be a poor choice of interpolation points for $\mathcal{H}_{2,\omega}$-MOR problem. Also, the accuracy in the tangential interpolation algorithms depends strongly on the choice of interpolation points and tangential directions. The final interpolation points and tangential directions of FLIRKA at convergence are not known a priori. FLIRKA may end up converging on the interpolation points and tangential directions which are worst for achieving less $\mathcal{H}_{2,\omega}$-error and vice versa. The fairness of comparison demands that most (if not all) of the interpolation points and tangential directions of FLPORK and FLIRKA are the same. Therefore, we have used six interpolation points and the tangential directions generated by FLIRKA and iterative rational Krylov algorithm (IRKA) \cite{gugercin2008h_2} (infinite frequency version of FLIRKA) at convergence in FLPORK and PORK, respectively, for a fair comparison. The desired frequency interval in FLBT, FLIRKA, FFMIRKA, and FLPORK is specified as $[0,8.93]$ rad/sec corresponding to the frequency of the mode $-0.25\pm j8.9339$. The frequency-domain error of the ROMs is shown in Fig. \ref{fig:3}.
It can be seen that FLPORK ensures a good accuracy within the desired frequency region where the most poorly damped modes lie. FLBT and FLIRKA are more accurate than FLPORK in the frequency-domain; however, they do not preserve the poorly damped modes of the original system. A $3$-phase fault is applied at bus $29$ of the study area at $0.1$ sec which is cleared at $0.2$ sec, and dynamic simulation is performed using the Power System Toolbox (PST) \cite{chow1992toolbox}. The time domain responses of the original system and the ROMs are shown in Fig. \ref{fig:4}. It can be seen that FLPORK also gives good accuracy in the time domain.
The time consumed by each algorithm to generate the ROM is shown in Table \ref{tab2a}. It can be noted that FLPORK is slightly more computational than PORK due to the computation of the integral $F(A)$ and $F(-S)$, but it is efficient as compared to FLBT and FLIRKA. Although SARQI and FFMIRKA took the most time here due to their iterative nature, FLBT is expected to become the most expensive as $n$ becomes greater than $2000$ due to the computation of dense large-scale Lyapunov equations. Next, we compare the simulation times of the experiments in Table \ref{tab3a}. It can be noted in Table \ref{tab3a} and Fig. \ref{fig:4} that the simulation time can significantly be reduced without a significant loss of accuracy.
\begin{table}[!h]
\centering
\caption{Comparison of the computational time}\label{tab2a}
\begin{tabular}{|c|c|}
\hline
Method             &  Time (sec)\\ \hline
SARQI & 6.17\\
PORK & 1.52\\
FLBT & 2.78\\
FLIRKA & 4.93\\
FFMIRKA & 10.96\\
FLPORK & 1.96\\\hline
\end{tabular}
\end{table}
\begin{table}[!h]
\centering
\caption{Comparison of the simulation time}\label{tab3a}
\begin{tabular}{|c|c|}
\hline
Method                &  Time (sec)\\ \hline
\textbf{Full Order}                &  12.95\\
SARQI & 5.01\\
PORK & 5.10\\
FLBT & 5.03\\
FLIRKA & 5.22\\
FFMIRKA & 6.36\\
FLPORK & 5.15\\\hline
\end{tabular}
\end{table}
\\\textit{\textbf{Test 2: IEEE 145-bus 50-machine system connected to NPCC 140-bus 40-machine system via Two Tie Lines:}} In this experiment, IEEE $145$-bus $50$-machine test system is considered as the external area, which is connected to the NPCC $140$-bus $40$-machine test system via two tie lines. The NPCC $140$-bus $40$-machine test system is the study area. The bus, line, and machine data of the external and study areas can be found in \cite{chow1992toolbox}. Bus-$93$ and bus-$104$ of the IEEE 145-bus 50-machine system are connected to the bus-$36$ and bus-$21$, respectively of the NPCC test system. The linear model for the external area is obtained according to equation (\ref{38a}), which is a $100^{th}$ order model with $4$ inputs and $2$ outputs. The inputs are the voltage magnitudes and angles on bus-$36$ and -$21$ of the NPCC test system, and the outputs are the rotor angles of generator-$1$ and -$2$ which are connected to bus-$93$ and -$104$ of the external area, respectively. The study area retains its nonlinear description. The rightmost poles of the external area are plotted in Fig. \ref{fig:5a}, and it can be seen that it has several poorly damped modes in its model.
 A $12^{th}$ order ROM of the external area is generated by SARQI -based modal truncation, PORK, FLIRKA, FLBT, FFMIRKA, and FLPORK. SARQI, PORK, FFMIRKA, and FLPORK preserve the most poorly damped four interarea and two local modes of the original system, i.e., $-0.2571\pm j3.936$ and $-0.2684\pm j4.765$, and $-0.1911\pm 11.37$, respectively to capture the interarea and local oscillations in the ROM. SARQI additionally preserves the following modes: $-0.2897\pm j5.097$, $-0.4143\pm j5.184$, and $-0.548\pm j7.7$ to make the ROM a $12^{th}$ order model. Again, we have used six interpolation points and the tangential directions generated by FLIRKA and IRKA \cite{gugercin2008h_2} in FLPORK and PORK, respectively, for a fair comparison. The desired frequency interval in FLBT, FLIRKA, FFMIRKA, and FLPORK is specified as $[0,11.37]$ rad/sec corresponding to the frequency of the mode $-0.1911\pm 11.37$. The frequency-domain error of the ROMs is shown in Fig. \ref{fig:6a}.
It can be seen that FLPORK ensures a good accuracy within the desired frequency region where the most poorly damped modes lie. A $3$-phase fault is applied at bus-$36$ of the study area at $0.1$ sec which is cleared at $0.5$ sec, and dynamic simulation is performed using PST \cite{chow1992toolbox}. The time domain responses of the original system and the ROMs are shown in Fig. \ref{fig:7a} and Fig. \ref{fig:8a}. It can be seen that FLPORK also gives good accuracy in the time domain as well.
The time consumed by each algorithm to generate the ROM is shown in Table \ref{tab4a}. Next, we compare the simulation times of the experiments in Table \ref{tab5a}. It can be noted in Table \ref{tab5a} and Fig. \ref{fig:7a}-\ref{fig:8a} that the simulation times can significantly be reduced without a significant loss of accuracy.
\begin{table}[!h]
\centering
\caption{Comparison of the computational time}\label{tab4a}
\begin{tabular}{|c|c|}
\hline
Method                 & Time (sec)\\ \hline
SARQI & 9.08\\
PORK & 1.95\\
FLBT & 3.59\\
FLIRKA & 8.19\\
FFMIRKA & 12.01\\
FLPORK & 2.14\\\hline
\end{tabular}
\end{table}
\begin{table}[!h]
\centering
\caption{Comparison of the simulation time}\label{tab5a}
\begin{tabular}{|c|c|}
\hline
Method                 &  Time (sec)\\ \hline
\textbf{Full Order}                & 17.05\\
SARQI & 6.29\\
PORK & 6.31\\
FLBT & 6.10\\
FLIRKA & 6.09\\
FFMIRKA & 7.17\\
FLPORK & 6.33\\\hline
\end{tabular}
\end{table}
\subsection{Reduced Order Damping Controller Design}
\begin{figure}[!h]
\centering
\includegraphics[width=8.4cm]{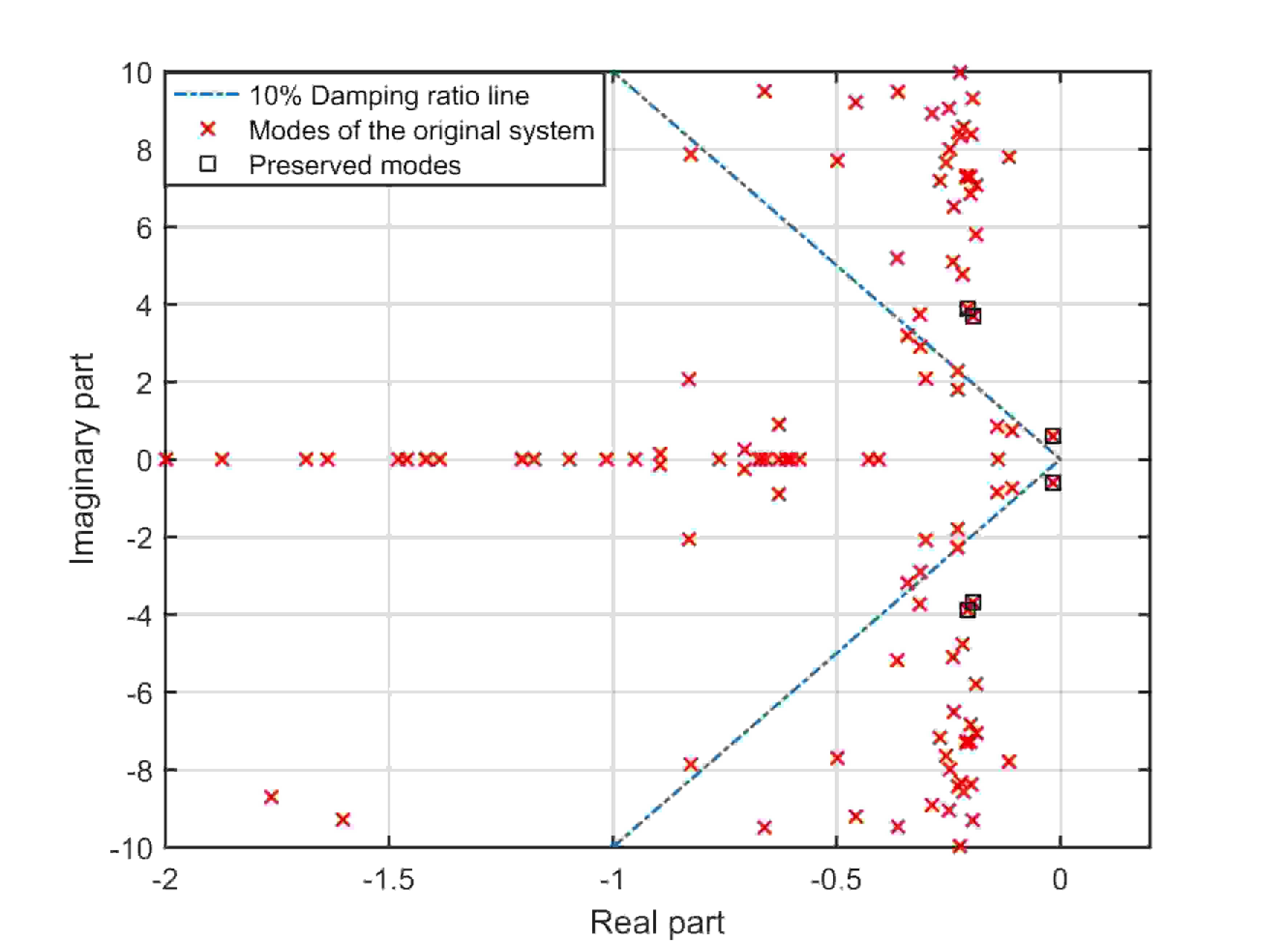}
\caption{The rightmost modes of $50$-machine test system}\label{fig:8}
\end{figure}

\begin{figure}[!h]
\centering
\includegraphics[width=8.4cm]{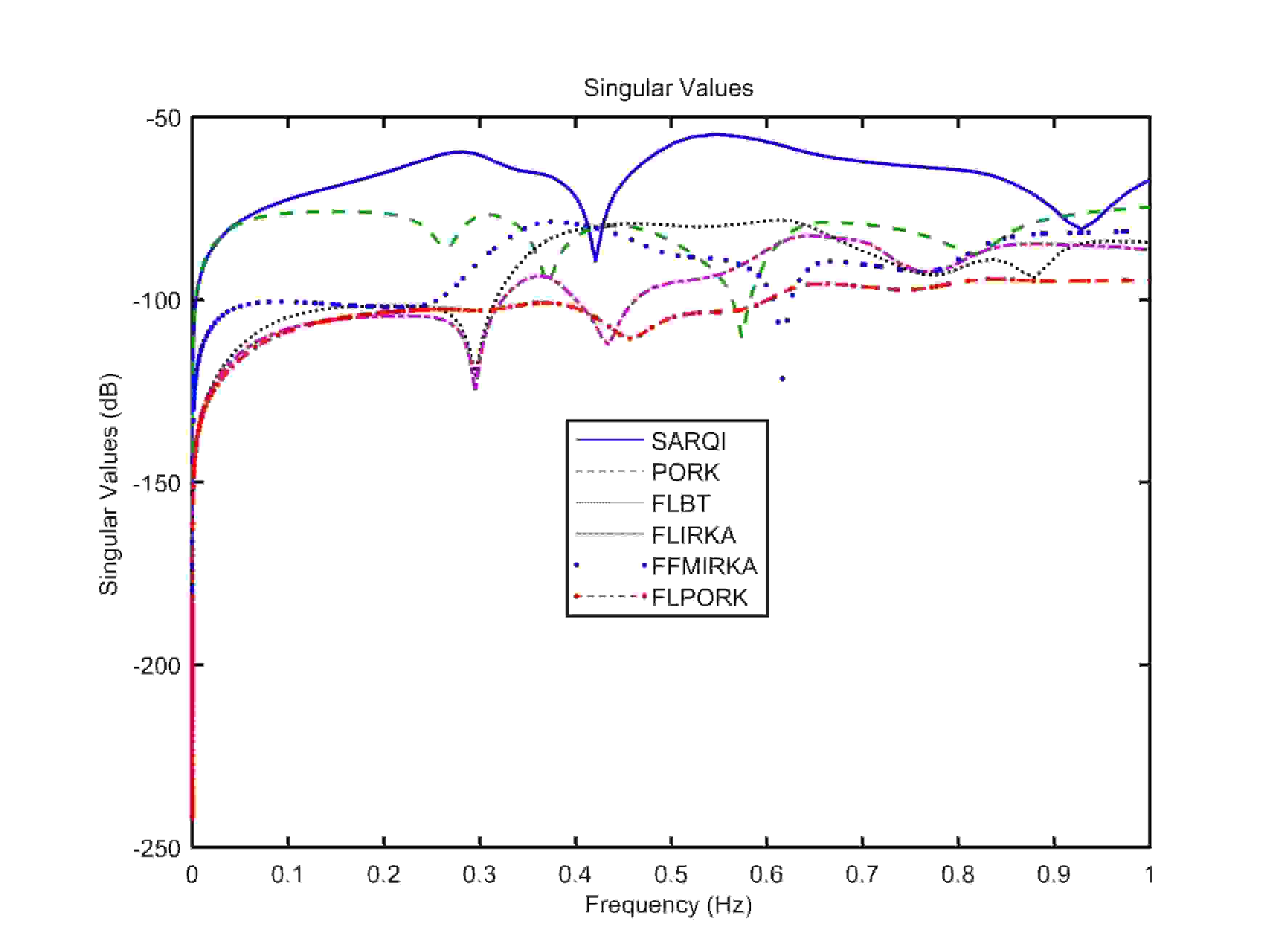}
\caption{Frequency domain error within $[0,1]$ Hz}\label{fig:9}
\end{figure}

\begin{figure}[!h]
\centering
\includegraphics[width=8.4cm]{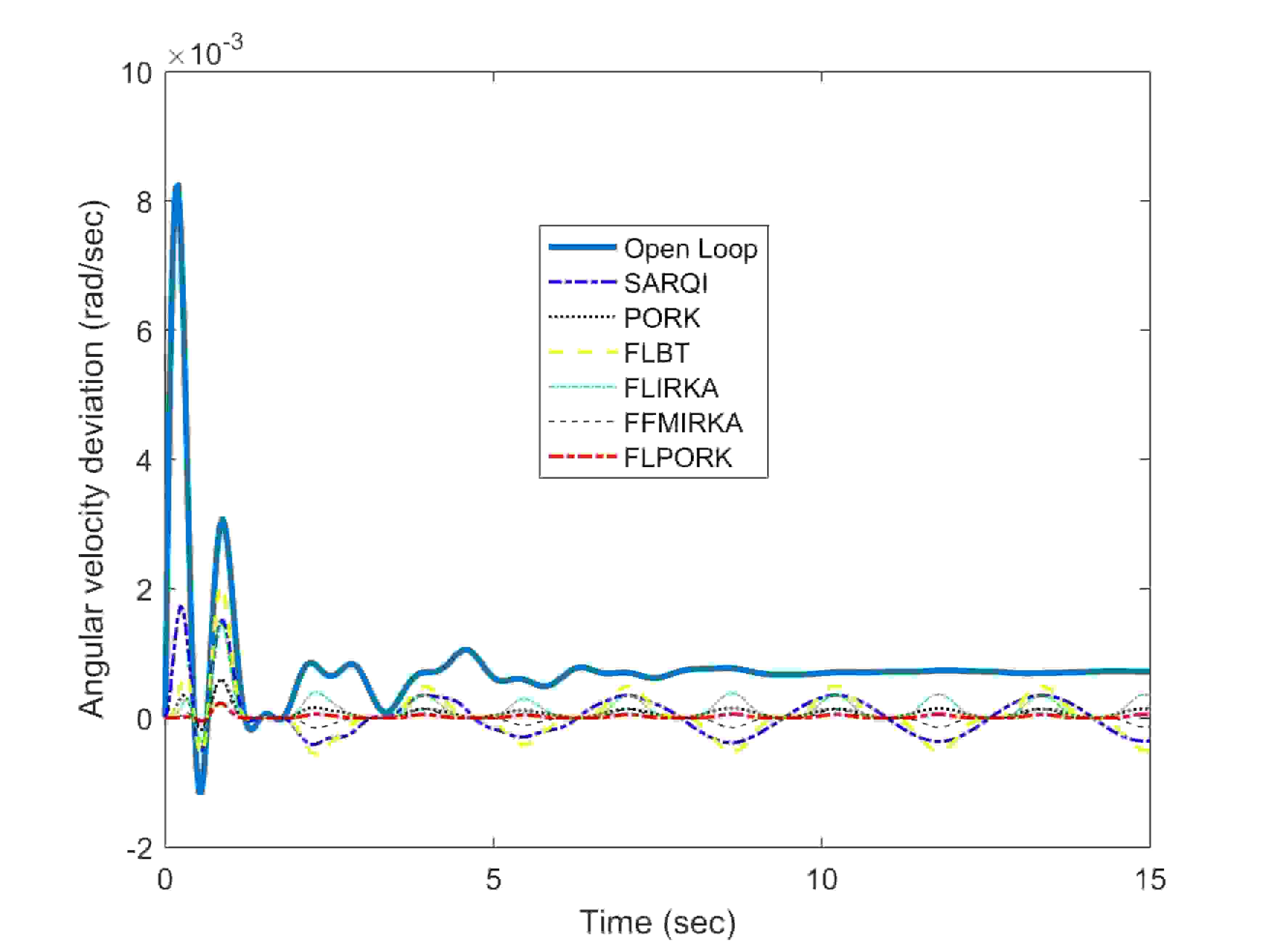}
\caption{Angular velocity deviation in G-21}\label{fig:10}
\end{figure}

\begin{figure}[!h]
\centering
\includegraphics[width=8.4cm]{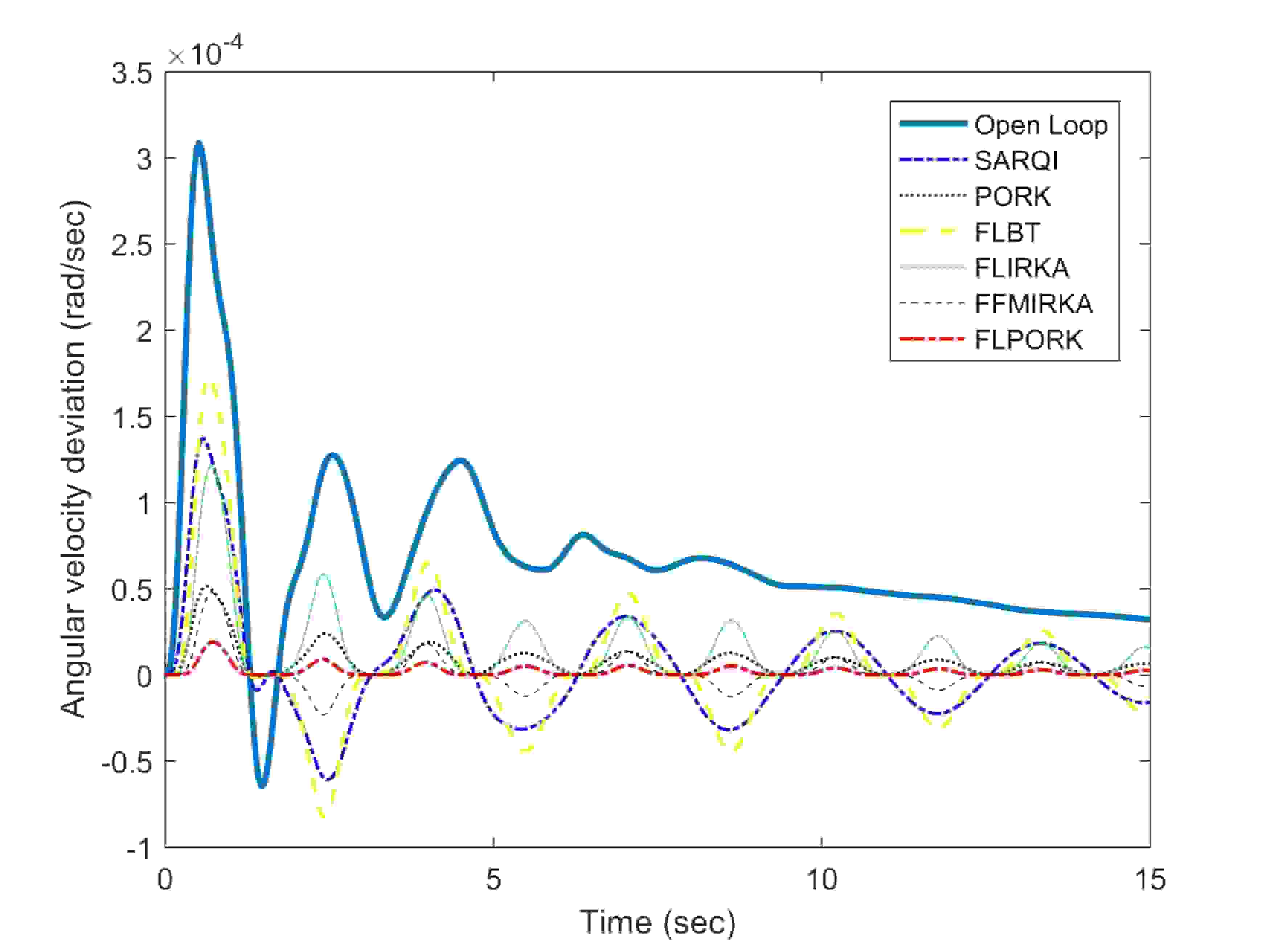}
\caption{Angular velocity deviation in G-24}\label{fig:11}
\end{figure}

\textit{\textbf{Test 3: Damping Controller for IEEE 145-bus 50-machine system:}} In this experiment, the reduced-order damping controller design problem for the IEEE $50$-machine $145$-bus system taken from \cite{chow1992toolbox} is considered. The linearized model of this test system is a $350^{th}$ order system obtained according to equations (\ref{39a})-(\ref{46a}). The four most poorly damped interarea modes are the following: $-0.2071\pm j3.8803$ and $-0.1945\pm j3.6916$. The important modes of the system are shown in Fig. \ref{fig:8}.
The $\mathcal{H}_\infty$ controller design yields a controller of the same order as that of the  plant which is impractical for the implementation. Therefore, we reduce the original model and then design the controller using the ROM. A $6^{th}$ order ROM is constructed using SARQI-based modal truncation, PORK, FLBT, FLIRKA, FFMIRKA, and FLPORK. SARQI, PORK, and FLPORK preserve the following modes of the original system: $-0.2071\pm j3.8803$, $-0.1945\pm j3.6916$, and $-0.0167\pm j0.5988$ while FFMIRKA preserves $-0.2071\pm j3.8803$ and $-0.1945\pm j3.6916$. The desired frequency interval is specified as $[0,3.88]$ rad/sec corresponding to the frequency of the mode $-0.2071\pm j3.8803$. Fig. \ref{fig:9} shows the frequency-domain error of the ROMs (corresponding to output-1). It can be noted that FLPORK ensures a good accuracy within the frequency region, which contains the electromechanical modes of the system.
The locations for the damping controller is identified using the participation factor method \cite{hsu1987identification}, and it suggests generator-$21$ (G-21) and generator-$24$ (G-24) be the appropriate locations. The angular velocity deviations from the reference angular velocity of these two generators are used as the feedback signals and the control input is added at $\Delta \bar{V}_{ref_i}$. A $6^{th}$ order decentralized $\mathcal{H}_\infty$ damping controller is designed for each ROM using the approach in \cite{furuya2006robust} which adds damping to the poorly damped interarea modes $-0.2071\pm j3.8803$ and $-0.1945\pm j3.6916$. A $10\%$ step change is induced in the mechanical torque of G-21 which in turn induces low-frequency oscillations. The angular velocity deviations in G-21 and G-24 with the open-loop and closed-loop systems are shown in Fig. \ref{fig:10} and Fig. \ref{fig:11}. It can be seen that the controller designed using the ROM generated by FLPORK provides maximum damping.
The interarea modes in the closed-loop systems are tabulated in Table \ref{tab1}.
\begin{table}[!h]
\centering
\caption{Interarea modes in the closed-loop systems}\label{tab1}
\begin{tabular}{|c|c|c|c|c|}
\hline
No                 & Method    & Mode & $\gamma$ $\%$ & $f$ $(Hz)$ \\ \hline
\multirow{7}{*}{1}        & \textbf{Open-loop}               & $-0.2071\pm j3.8803$& $5.32$& $0.6176$\\
                          & SARQI                   & $-0.3113\pm j3.9336$& $7.89$& $0.6261$\\
                          & PORK                    & $-0.3060\pm j3.8628$ &  $9.01$& $0.6148$\\
                          & FLBT                & $-0.3071\pm j3.9404$ & $7.77$& $0.6271$\\
                          & FLIRKA                & $-0.3123\pm j3.8380$ & $8.11$& $0.6108$\\
                          & FFMIRKA                & $-0.3539\pm j3.9291$ & $8.97$& $0.6253$\\
                          & FLPORK               & $-0.4369\pm j3.8662$ &  $11.23$&$0.6153$\\ \hline
\multirow{7}{*}{2}        & \textbf{Open-loop}               & $-0.1945\pm j3.6916$& $5.26$& $0.5875$\\
                          & SARQI                   & $-0.2582\pm j3.5178$& $7.32$& $0.5599$\\
                          & PORK                    & $-0.2886\pm j3.4656$ &  $8.30$& $0.5516$\\
                          & FLBT                & $-0.2319\pm j3.7398$ & $6.19$& $0.5952$\\
                          & FLIRKA                & $-0.3483\pm j3.6382$ & $9.53$& $0.5790$\\
                          & FFMIRKA                & $-0.3596\pm j3.6744$ & $9.74$& $0.5848$\\
                          & FLPORK               & $-0.3803\pm j3.4302$ &  $11.07$&$0.5459$\\ \hline
\end{tabular}
\end{table}
\section{Conclusion}\label{Sec:V}
In this paper, a frequency-limited MOR technique is proposed which yields a ROM which not only satisfies a subset of the first-order optimality conditions of the problem $||G(j\omega)-\tilde{G}(j\omega)||^2_{\mathcal{H}_{2,\omega}}$ but also preserves the electromechanical modes of the power system. The proposed algorithm can generate an accurate ROM with the desired modes, which ensures a good accuracy in the desired frequency interval. The proposed algorithm is applicable to large-scale systems and hence can be used for fast dynamic simulation and reduced order controller design for large-scale power systems.
\section*{Acknowledgment}
The first author would like to thank M. A. Pai of the University of Illinois at Urbana-Champaign, Urbana, USA for explaining his work \cite{chaniotis2005model,liu2013krylov}, and answering several questions related to power system modeling for MOR done in his students' work \cite{liu2009dynamic,chaniotis2001krylov}. This work was supported by National Natural Science Foundation of China under Grant (No. $61873336$, $61873335$), and supported in part by $111$ Project (No. D$18003$).
\ifCLASSOPTIONcaptionsoff
  \newpage
\fi

\end{document}